\RequirePackage[2020-02-02]{latexrelease}
\documentclass{aa}  

\usepackage{aas_macros}

\RequirePackage[utf8]{inputenc} 
\RequirePackage[T1]{fontenc}

\DeclareUnicodeCharacter{03BC}{$\mu$}
\DeclareUnicodeCharacter{B5}{$\mu$}
\DeclareUnicodeCharacter{0394}{$\Delta$}

\usepackage{tabulary}
\usepackage{lscape}
\usepackage{caption}
\RequirePackage{longtable}

\PassOptionsToPackage{dvipsnames}{xcolor} 
\usepackage{pgf}
\usepackage[inline]{enumitem}

\usepackage{graphicx,txfonts,array}
\usepackage{url}
\usepackage{xspace}

\usepackage[switch]{lineno}
\usepackage{natbib}
\bibpunct{(}{)}{;}{a}{,}{,}

\usepackage{color}
\graphicspath{{figures/}}

\usepackage{tipa}

\usepackage[colorlinks=True,allcolors=blue]{hyperref}

\usepackage[nameinlink]{cleveref}

\usepackage{catoptions}

\makeatletter

\def\Autoref#1{%
  \begingroup
  \edef\reserved@a{\cpttrimspaces{#1}}%
  \ifcsndefTF{r@#1}{%
    \xaftercsname{\expandafter\testreftype\@fourthoffive}
      {r@\reserved@a}.\protect\\{#1}%
  }{%
    \ref{#1}%
  }%
  \endgroup
}
\def\testreftype#1.#2\\#3{%
  \ifcsndefTF{#1autorefname}{%
    \def\reserved@a##1##2\@nil{%
      \uppercase{\def\ref@name{##1}}%
      \csn@edef{#1autorefname}{\ref@name##2}%
      \autoref{#3}%
    }%
    \reserved@a#1\@nil
  }{%
    \autoref{#3}%
  }%
}

\newcommand{\nuna}[2]{(#1)\,\textit{#2}} 

\newcommand{\numb}[1]{\textcolor{orange}{#1}}
\renewcommand{\numb}[1]{#1}

\newcolumntype{L}{>{\raggedright\arraybackslash}p{3.7cm}}
\newcolumntype{M}{>{\raggedright\arraybackslash}p{4.5cm}}
\newcolumntype{N}{>{\raggedright\arraybackslash}p{5.0cm}}
\newcolumntype{P}{>{\raggedright\arraybackslash}p{9.5cm}}
\newcolumntype{Q}{>{\raggedright\arraybackslash}p{12cm}}
\newcolumntype{R}{>{\raggedright\arraybackslash}p{7.0cm}}

\newcommand{\miriade}{\texttt{Miriade}\xspace}
\newcommand{\skybot}{\texttt{SkyBoT}\xspace}
\newcommand{\ssodnet}{\texttt{SsODNet}\xspace}
\newcommand{\elasticsearch}{Elasticsearch\xspace}

\newcommand{\quaero}{\texttt{quaero}\xspace}
\newcommand{\datacloud}{\texttt{dataCloud}\xspace}
\newcommand{\ssocard}{\texttt{ssoCard}\xspace}
\newcommand{\bft}{\texttt{ssoBFT}\xspace}
\newcommand{\rocks}{\texttt{rocks}\xspace}

\newcommand{\mpcorb}{\texttt{mpcorb}\xspace}
\newcommand{\astorb}{\texttt{astorb}\xspace}
\newcommand{\cometpro}{\texttt{cometpro}\xspace}

\newcommand{\diamalbedo}{\texttt{diamalbedo}\xspace}
\newcommand{\masses}{\texttt{masses}\xspace}
\newcommand{\density}{\texttt{density}\xspace}
\newcommand{\spin}{\texttt{spin}\xspace}
\newcommand{\taxonomy}{\texttt{taxonomy}\xspace}
\newcommand{\colors}{\texttt{colors}\xspace}
\newcommand{\thermal}{\texttt{thermal\_properties}\xspace}
\newcommand{\mpcatobs}{\texttt{mpcatobs}\xspace}
\newcommand{\phase}{\texttt{phase\_function}\xspace}
\newcommand{\proper}{\texttt{proper\_elements}\xspace}
\newcommand{\family}{\texttt{family}\xspace}
\newcommand{\families}{\texttt{families}\xspace}
\newcommand{\yarko}{\texttt{yarkovsky}\xspace}
\newcommand{\pairs}{\texttt{pairs}\xspace}

\newcommand{\urlRocks}{https://github.com/maxmahlke/rocks}
\newcommand{\urlRocksDocs}{https://rocks.readthedocs.io}

\newcommand\nbssonames{$1\,288\,838$\xspace}  
\newcommand\nbssodesignations{$5\,360\,208$\xspace}  

\newcommand\datacloudnbcollections{15\xspace}  
\newcommand\datacloudnbfields{591\xspace}  
\newcommand\datacloudnbmaxcolumns{651\xspace}  

\newcommand\bftfilesize{2.1\xspace}  
\newcommand\bftnbrows{1\,223\,984\xspace}  
\newcommand\bftfilledfields{105\xspace}  
\newcommand\bftfillingfactor{14.6\%\xspace}  
\newcommand\bftfillingfactorapprox{$\sim$15\%\xspace}  

\newcommand\datacloudrefdatasets{3007\xspace}  

\newcommand\colorsNBibRef{29\xspace} 
\newcommand\diamalbedoNBibRef{205\xspace} 
\newcommand\taxonomyNBibRef{208\xspace} 
\newcommand\massesNBibRef{165\xspace} 
\newcommand\phasefunctionNBibRef{4\xspace} 
\newcommand\familiesNBibRef{9\xspace} 
\newcommand\pairsNBibRef{12\xspace} 
\newcommand\spinNBibRef{$2\,775$\xspace} 
\newcommand\yarkovskyNBibRef{17\xspace} 
\newcommand\thermalpropertiesNBibRef{57\xspace} 
\newcommand\densityNBibRef{26\xspace} 
\newcommand\astorbN{$1\,078\,203$\xspace} 
\newcommand\astorbNSSO{$1\,078\,203$\xspace} 
\newcommand\mpcorbN{$1\,223\,386$\xspace} 
\newcommand\mpcorbNSSO{$1\,223\,386$\xspace} 
\newcommand\cometproN{$1\,613$\xspace} 
\newcommand\cometproNSSO{$1\,613$\xspace} 
\newcommand\mpcatobsN{$\,341\,772\,068$\xspace} 
\newcommand\mpcatobsNSSO{$1\,674\,187$\xspace} 
\newcommand\colorsN{$4\,793\,938$\xspace} 
\newcommand\colorsNSSO{$\,428\,339$\xspace} 
\newcommand\densityN{49\xspace} 
\newcommand\densityNSSO{29\xspace} 
\newcommand\diamalbedoN{$\,261\,396$\xspace} 
\newcommand\diamalbedoNSSO{$\,149\,375$\xspace} 
\newcommand\familiesN{$\,493\,364$\xspace} 
\newcommand\familiesNSSO{$\,261\,832$\xspace} 
\newcommand\massesN{$2\,170$\xspace} 
\newcommand\massesNSSO{422\xspace} 
\newcommand\pairsN{340\xspace} 
\newcommand\pairsNSSO{236\xspace} 
\newcommand\phasefunctionN{$\,330\,279$\xspace} 
\newcommand\phasefunctionNSSO{$\,227\,888$\xspace} 
\newcommand\properelementsN{$\,799\,878$\xspace} 
\newcommand\properelementsNSSO{$\,799\,878$\xspace} 
\newcommand\spinN{$47\,541$\xspace} 
\newcommand\spinNSSO{$28\,951$\xspace} 
\newcommand\taxonomyN{$\,274\,322$\xspace} 
\newcommand\taxonomyNSSO{$\,140\,713$\xspace} 
\newcommand\thermalinertiaN{$4\,510$\xspace} 
\newcommand\thermalinertiaNSSO{$2\,109$\xspace} 
\newcommand\yarkovskyN{826\xspace} 
\newcommand\yarkovskyNSSO{578\xspace} 

\begin{document} 

   \title{\ssodnet: Solar system Open Database Network\thanks{%
     The \bft catalog is available
     at the CDS via anonymous ftp to
     \url{http://cdsarc.u-strasbg.fr/} or via
     \url{http://cdsarc.u-strasbg.fr/viz-bin/qcat?J/A+A/xxx/Axxx}}}

   \author{
     J.~Berthier\inst{\ref{imcce}}   \and
     B.~Carry\inst{\ref{oca}}        \and
     M.~Mahlke\inst{\ref{oca}}       \and
     J.~Normand\inst{\ref{imcce}}    }

   \institute{
     IMCCE, Observatoire de Paris, PSL Research University, CNRS, Sorbonne Universit{\'e}s, UPMC Univ Paris 06, Univ. Lille, France\\
     \email{jerome.berthier@obspm.fr, jonathan.normand@obspm.fr}\label{imcce}
     \and
     Universit\'e C{\^o}te d'Azur, Observatoire de la
     C{\^o}te d'Azur, CNRS, Laboratoire Lagrange, France\\
     \email{benoit.carry@oca.eu, max.mahlke@oca.eu}\label{oca}
   }

   \date{.../...}

 
  \abstract
   {The sample of Solar system objects has 
    dramatically increased over the last decade.
    The number of measured properties (e.g., diameter, taxonomy, rotation period, 
    thermal inertia, etc.) has expanded even more quickly. However, this wealth of information
    is spread over a myriad of studies, with different designations reported per object.}
   {We provide a solution to 
    the identification of Solar system objects 
    based on any of their multiple names or designations. We also
    compile and rationalize 
    their properties to provide an easy access to them.
    We aim to continuously update the database as new measurements become available.}
   {We built a Web Service, \ssodnet, which offers four access points,
    each corresponding
    to an identified necessity in the community: 
    name resolution (\quaero), 
    compilation of a large corpus of properties (\datacloud),
    determination of the best estimate among compiled values (\ssocard),
    and a statistical description of the population (\bft).}
   {The \ssodnet interfaces are fully operational and freely accessible to everyone.
    The name resolver \quaero translates any of the $\sim$\numb{5.3\ million} designations 
    of objects into their current and official designation.
    The \datacloud includes about \numb{\bftfilledfields\ million} parameters
    (osculating and proper elements, pair and family membership, 
    diameter, albedo, mass, density, rotation period, spin coordinates, phase function 
    parameters, colors, taxonomy, thermal inertia, and Yarkovsky drift)
    from over \numb{3,000} articles (updated continuously).
    For each of the known asteroids and dwarf planets ($\sim$\numb{1.2\ million}), 
    a \ssocard that provides a single best-estimate for  each parameter is available.
    The \ssodnet service provides these resources in a fraction of second upon query.
    Finally, the extensive \bft table compiles all the best estimates
    in a single table for population-wide studies.}
   {}

   \keywords{Astronomical data bases --
      Catalogs -- 
      Minor planets, asteroids: general
   }

   \maketitle
%

%

\section{Introduction}

  The first decade of the 2000s saw an order of magnitude increase in the sample
  of known Solar System objects (SSOs) from roughly \numb{50,000} 
  to \numb{600,000}.
  While this number has doubled since, the revolution of most recent decade
  has seen an even faster growth on the part of the measured properties of these bodies.
  About \numb{2000} diameters and albedo had been determined from IRAS
  mid-infrared observations \citep{2002AJ....123.1056T} and
  over \numb{150,000} are available today
  \citep[e.g.,][]{2011ApJ...743..156M, 2011ApJ...741...68M, 2011ApJ...742...40G}.
  Hundreds of detections of the Yarkovsky effect \citep{2015-AsteroidsIV-Vokrouhlicky}
  are available \citep[e.g.,][]{2019A&A...627L..11D, 2020AJ....159...92G} just 20 years after the first-ever detection \citep{2003Sci...302.1739C}.

  This wealth of characterizations (e.g., colors, albedos, rotation periods, etc.) has allowed for
  multiple statistical studies on the forced orientation of family members by the 
  Yarkovsky-O'Keefe-Radzievskii-Paddack (YORP) effect
  \citep{2002Natur.419...49S, 2016A&A...586A.108H}, 
  the compositional distribution of the asteroid belt 
  \citep{2014Natur.505..629D}, 
  the size-frequency distribution of asteroid families
  \citep{2008Icar..198..138P,2013ApJ...770....7M}, 
  the internal structure of minor bodies
  \citep{2012PSS...73...98C,2015-AsteroidsIV-Scheeres}, 
  and the origins of near-Earth asteroids
  \citep{2018P&SS..157...82P,2019AJ....158..196D,2019Icar..324...41B}, 
  among many others.

  The benefit of all these developments has not, however, come to full fruition.
  If some catalogs are publicly available in machine-readable formats on the
  Planetary Data System\footnote{\url{https://pds.nasa.gov/}} (PDS), 
  the Centre de Données astronomiques de Strasbourg\footnote{\url{https://cdsweb.u-strasbg.fr/}} (CDS), 
  or alternative repositories (with unfortunately an endless variety of formats),
  a significant fraction of results have only been tabulated within the relevant papers.
  Some journals offer machine-readable versions of these tables on their online versions,
  but only for recent articles.
  Furthermore, the designation of small bodies often evolves over time, 
  going from several possible provisional designations to a single number and then ultimately
establishing  an official name. Hence, the same object can be referred to with different
  labels in different studies, making its cross-identification over several
  sources a complex task.
  Accessing to all the characteristics of a given body or population
  can thus prove tedious and even impractical.

  Compiling estimates of SSO properties and deriving the best estimate for each
  is of high practical relevance for the computation of ephemerides in 
  the Virtual Observatory (VO) Web services we maintain
  \citep[\miriade, \skybot,][]{2006ASPC..351..367B,2008LPICo1405.8374B}.
  Dynamical properties (i.e., osculating elements) are required to compute the position
  of SSOs and physical properties are required to predict their apparent aspect
  as seen by an observer, such as 
  \begin{enumerate*}[label=(\roman*)]
    \item the apparent magnitude in V band, relying on the phase function
      \citep[HG or HG$_1$G$_2$,][]{1989-AsteroidsII-Bowell, 2010Icar..209..542M},
    \item the apparent magnitude in any other band, requiring a color index derived
        from the spectral class \citep[e.g.,][]{2013Icar..226..723D,2018A&A...617A..12P},
    \item the flux at mid-infrared wavelengths, computed from the diameter and albedo
        through a thermal model \citep{1999Icar..142..464H},
    \item the shape and orientation of a target on the plane of the sky
        (often referred to as physical ephemerides), based on its
        3D shape model, rotation period, and spin-vector coordinates
        \citep[e.g.,][]{2012A&A...545A.131M}.
  \end{enumerate*}

  Beyond the ephemerides computation, an extensive and rationalized compilation 
  of SSO properties has many applications, from detailed in-depth
  studies on specific targets to population-wide statistical description 
  of parameters.
  Over the years, publicly available compilations of data 
  have flourished, for instance,
  the Jet Propulsion Laboratory Small Bodies Database\footnote{\url{https://ssd.jpl.nasa.gov/}},
  the Las Cumbres Observatory NEOExchange\footnote{\url{https://neoexchange.lco.global/}} \citep{2021Icar..36414387L}, 
  the Lowell observatory Minor Planet Services\footnote{\url{https://asteroid.lowell.edu}}
  \citep{2021DPS....5310104M},
  the NEOROCKs physical properties database\footnote{\url{https://neorocks.elecnor-deimos.com/web/guest/search-retrieval}}
  \citep{2021LPICo2549.7032Z},
  the Observatoire de la Côte d'Azur Minor Planet Physical Properties Catalog\footnote{\url{https://mp3c.oca.eu}} \citep[MP3C,][]{2018-ACM-Delbo},
  the Size, Mass and Density of Asteroids \citep[SiMDA\footnote{\url{https://astro.kretlow.de/?SiMDA}},][]{2020EPSC...14..690K},
  and
  the SUPAERO ECOCEL\footnote{\url{http://www.ecocel-database.com}} \citep{2022P&SS..21505463K}.
  While these services fulfill many of the community's needs, most of them do not provide
  a fast machine-machine interface.
 
  Thus, we have designed a fully scriptable Web Service named 
  the  Solar system Open Database Network\ (\ssodnet) which is aimed at providing
  the best estimate of a variety of parameters for every SSO.
  Owing to the complexity of compiling SSO data as depicted hereinabove, \ssodnet 
  consists of a suite of chained steps: from the identification of objects to
  the massive compilation of data, ending with the selection of best estimates, 
  and summarizing them in a table.
  As each of these steps represents a typical task relevant for the community, we 
  propose a dedicated front-end (a Web service associated with an
  Application Programming Interface - API) for each.

  In \Autoref{sec:quaero}, we describe how \quaero builds a unique identifier for each object, 
  associating all its aliases and providing the identity of the SSO. 
  In \Autoref{sec:datacloud}, we describe how \datacloud compiles the measurements and 
  estimates of properties from many sources, providing the most-possible comprehensive 
  data set of SSOs.
  In \Autoref{sec:ssocard}, we describe how \ssocard provides the best estimate of each 
  SSO property, and lists them in a single organized identify card.
  In \Autoref{sec:bft}, we describe how \bft summarizes the most-commonly requested 
  of these parameters for all SSOs.
  We then describe how to query these services in \Autoref{sec:access}
  before discussing the future developments of \ssodnet in \Autoref{sec:future}.

\section{Name resolver: \quaero\label{sec:quaero}}

  The {\ssodnet}.{\quaero} name-resolution service is built to address the 
  issue of identification of SSOs and, more generally, of all planetary and
  artificial objects gravitationally bound to a star. Upon the submission 
  of any of the possible designations of a target, \quaero returns its official or 
  main designation, together with all its aliases. To be compliant with 
  the spirit of the VO (``name resolver'') \quaero can also return the equatorial coordinates
  of the object at a given epoch.

  \subsection{Context}

    The Solar System is populated by widely different types of celestial bodies: 
    from planets and their satellites to minor planets (comets, asteroids, Centaurs, 
    Kuiper-belt objects, etc.) and their satellites, and further on to artificial satellites, space probes, 
    and space debris. Since the first exoplanet detection by
    \citet{1995Natur.378..355M},
    today we know of about 5000 planetary objects that orbit around 
    other stars than the Sun. A few rogue planets (e.g., OTS 44 or Cha 110913-773444) 
    and two interstellar objects (1I/$^\prime$Oumuamua and 2I/Borisov) complete the 
    picture of the planetary zoo.

    The nomenclature of SSOs is entrusted to two groups under the auspices of the
    International Astronomical Union (IAU) Division F. The first is the Working Group 
    for Planetary System Nomenclature (WG-PSN), which is in charge of naming features on 
    planets, satellites, and asteroids. This group also names planets (although the 
    IAU has not named a planet as of yet) and the natural satellites of major planets.
    The second is the Working Group for Small Bodies Nomenclature (WG-SBN), which is responsible 
    for naming of all other small bodies (minor planets, satellites of minor planets, 
    and comets). Both working groups share responsibility for naming dwarf planets 
    \citep{IAU-naming-web-site}.
    
    As of today, there are no official name for exoplanets assigned by the IAU.
    The public names, assigned through a public naming process such as 
    NameExoWorlds\footnote{\url{https://www.iau-100.org/name-exoworlds}}
    is distinguished from the official scientific designation, which follows the
    rules of the system used for designating multiple-star systems as adopted by
    the IAU \citep{IAU-exoplanets-web-site}.

    Spacecraft, together with launchers, payloads, and space debris, are indexed for 
    safety and cooperation purposes. They are usually named by their funders (space 
    agencies, laboratories, or companies). They are also assigned an International 
    Designator (COSPAR ID), under the responsibility of the \emph{Committee on Space 
    Research} (COSPAR) of the \emph{International Council for Science} (ICSU), and a 
    Satellite Catalog Number (NORAD ID) attributed by the \emph{United States Space 
    Command} (USSPACECOM).
    
\begin{figure}[t]
      \centering
      \input{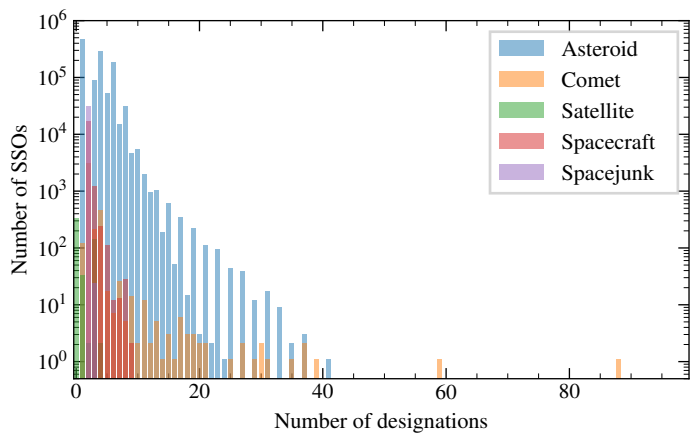}
      \caption{Histogram of the number of designations for each class of object.}
      \label{fig:names}
    \end{figure}

    Since the designation of the major bodies of the Solar System (the Sun, the Moon, 
    the eight planets), more than \numb{1.2 million} objects have been inventoried, 
    classified, and named. As of today, there are more than \numb{5.3} million designations
    used to name them all. Objects can have multiple designations owing to the 
    evolution of knowledge as well as changes in nomenclature over
    time\footnote{\url{https://www.minorplanetcenter.net/iau/info/DesDoc.html}}.
    We illustrate this with the first asteroid discovered in 1801: Ceres.
    It is classified today as a dwarf planet.
    Its official designation is \textquotedblleft (1) Ceres:\textquotedblright\ 
    a number in parenthesis followed by a name. This official designation thus 
    already contains two labels. However, Ceres was also named using provisional 
    designations over the years, assigned to past astrometric observations that 
    had not been immediately connected to its orbit: 
    \textquotedblleft 1801 AA,\textquotedblright\ 
    \textquotedblleft 1899 OF,\textquotedblright\ 
    and \textquotedblleft 1943 XB,\textquotedblright\  
    and the corresponding packed
    names\footnote{\url{https://www.minorplanetcenter.net/iau/info/PackedDes.html}}
    \textquotedblleft I01A00A,\textquotedblright\ 
    \textquotedblleft I99O00F,\textquotedblright\ 
    and \textquotedblleft J43X00B\textquotedblright. 
    Thus Ceres can be known by eight different names.
    The all-time record is held by comets P/Halley (1P), with 59 designations,
    and P/Encke (2P), with 89 designations.
    We present in \Autoref{fig:names} the distribution of the number of
    designations by type of SSOs and we present a summary in \Autoref{tab:names}.

    \begin{table}[t]
      \centering
      \caption{Statistics of the number of SSO designations by object class.}
      \label{tab:names}
      \begin{tabular}{lcccc}
        \hline 
        \multicolumn{1}{c}{Type} & \multicolumn{4}{c}{Number of designations}\\ 
          & min & max & mean & $\sigma$ \\
        \hline 
        Asteroids     & 2 & 42 & 4 & 2 \\
        Comets        & 2 & 89 & 4 & 3 \\
        Dwarf planets & 6 & 10 & 6 & 2 \\
        Planets       & 3 &  3 & 3 & 0 \\
        Satellites    & 2 &  6 & 2 & 1 \\
        Spacecrafts   & 3 & 10 & 3 & 1 \\
        Spacejunks    & 3 &  4 & 3 & 0 \\
        Exoplanets    & 1 &  2 & 1 & 0 \\
        \hline 
      \end{tabular} 
    \end{table}

\subsection{Quaero: \normalfont\textsf{\textipa{/"k\textsuperscript{w}a\u*{e}.ro:/}}}

   This is the core of \ssodnet. It ensures the reliability of the naming
   of SSOs and it allows us to cross-match identifications between their actual 
   names and the designations used over time in the various data sets. 
   In \numb{August 2022}, we counted \numb{\nbssonames} solar and extra-solar objects 
   for \numb{\nbssodesignations} designations (1:4 ratio).

   Overall, \ssodnet.\quaero is designed to fulfill four main functionalities:
   1) to identify a SSO from its designation; 
   2) to explore the naming of SSOs using wildcard, regular expression, or fuzziness; 
   3) to resolve the name of a SSO into sky coordinates; and 
   4) to provide an autocomplete feature that can be used to offer SSO name suggestions 
   when a user types in an input field.

   To achieve this goal, once a week we gather all the available planetary object designations 
   from the Minor Planet Center \citep{1980CeMec..22...63M} for asteroids and dwarf planets,
   the IMCCE's CometPro Database \citep{1996IAUS..172..357R} for comets,
   the Extrasolar Planets Encyclopaedia \citep{exoplanet.eu} for exoplanets,
   and CelesTrak \citep{celestrack} for spacecrafts and debris.
   These designations are then stored and indexed in a dedicated database.

   We use the NoSQL database \elasticsearch\footnote{\url{https://www.elastic.co/}} to 
   manage the millions of designations. It is a full-text search engine based on the 
   Apache Lucene library\footnote{\url{https://lucene.apache.org/}}. Each object 
   is defined by a set of fields (document) defining its Id, name, aliases, parent, type, 
   and so on. Documents are stored in an \elasticsearch index as JSON-format data. By 
   default, \elasticsearch tries to guess the correct mapping for fields, but to meet the 
   challenges of planetary object identification, we specified our own mapping.

   If SSO designations are indexed as individual strings, then a user can only find 
   whole names. To allow for the search of a name on a part of a designation, we decompose all 
   the SSO designations into small chunks (tokens). However, at this step, each token is still 
   matched literally. This means (among other things) that a search for a name with or without
   an accent or a special character, or one with mixed lowercase and uppercase characters, 
   would possibly not result in a match with any name. To solve this issue, we defined the 
   normalization rules to allow for the matching of tokens that are not exactly the same as 
   the search names, but similar enough to still be relevant.
   For the full technical information, we refer to the documention\footnote{\url{https://doc.ssodnet.imcce.fr/quaero.html}}
   of the \ssodnet.\quaero API.

\section{Compilation of properties: \datacloud\label{sec:datacloud}}

  The {\ssodnet}.{\datacloud} service is designed to compile all published 
  measurements and estimates of SSO properties. The \datacloud uses
  \ssodnet.\quaero to identify objects over their multiple designations.
  It also associates every estimate with a bibliographic reference
  and a method. Upon request, the \datacloud returns all the estimates 
  of a given property or parameter for the requested SSO.

  \subsection{Context}

    Starting with the planetary motion \citep{1760pnpm.book.....N}, 
    the first studies of SSOs focused on their dynamics
    \citep{1809tmcc.book.....G}, which is required to compute their ephemerides.
    From the distribution of their orbital elements, 
    \citet{1918AJ.....31..185H} discovered the dynamical families.
    Time-series photometry has led to the determination of numerous rotation
    periods in the first half of the twentieth century
    \citep[e.g.,][]{1913AnHar..72..165B}.
    The 1970s saw the advent of compositional and physical studies,
    with the first studies of diameter and albedo 
    \citep[e.g.,][]{1973Icar...20..477C}, mass and, hence, the density 
    \citep{1974A&A....30..289S}, along with the spectrophotometry and taxonomy 
    \citep[e.g.,][]{1975Icar...25..104C}.
    The handful of SSOs with spin-vector coordinates and
    triaxial dimensions of the 1980s \citep{1989Icar...78..323D} grew to several
    hundreds in the 2000s thanks to the light-curve inversion technique
    \citep{2001Icar..153...24K}.
    Similarly, estimates of thermal inertia 
    and Yarkovsky drift are common nowadays 
    \citep{2018Icar..309..297H, 2020AJ....159...92G}, even though the first studies
    were completed only two decades ago
    \citep{1996A&A...310.1011L, 2003Sci...302.1739C}.

    Benefiting from these progresses is complex, however, as the
    fast-growing number of measured properties
    is spread over a myriad of articles.
    Machine-readable catalogs delivered by authors to 
    the PDS or the CDS only represent the tip of the iceberg.
    Furthermore, there is a large heterogeneity in how SSOs are
    labeled (number, name, packed designation, etc.) and in how
    quantities are reported:
    masses, $M,$ in terms of kg or solar masses (M$_\odot$) or as 
    a GM product or the albedo in linear or logarithmic scale,
    for instance.
    
    The sample size of individual articles may be small, but their sum 
    is large. In particular, some size-limited sample may be extremely 
    valuable, such as results on a single target obtained during
    a spacecraft rendezvous for example. Therefore, the goal of compiling every
    estimate should not be overlooked by the community.

    \ssodnet.\datacloud compiles in a single database as many estimates
    as possible for a variety of SSO properties.
    Such a centralization of data may appear anachronistic in the current landscape of
    connections to remote databases, such as what is regularly done in the VO
    \citep{2008A&A...492..277B}. It is, however, required here. 
    First, the remote databases do not exist. Second, owing to the
    issue of SSO naming, on-the-fly cross-matches between resources
    would be slow upon query.
    We chose to place the workload on the server side, in an asynchronous
    process, to provide a fast service to users.
    Such a solution is already used for the ESA Gaia
    archive\footnote{\url{https://gea.esac.esa.int/archive/}}, in
    which time-consuming cross-matches of Gaia catalog 
    \citep{2016A&A...595A...2G, 2018A&A...616A...1G, 2021A&A...649A...1G}
    with other common large catalogs
    \citep[e.g., SDSS DR9, 2MASS, allWISE,][]{%
      2012ApJS..203...21A, 2006AJ....131.1163S, 
      2010AJ....140.1868W, 2013wise.rept....1C}
    are already computed and stored
    \citep[see details in][]{2017A&A...607A.105M}.

  \subsection{Method\label{ssec:datacloud:method}}

    The design of the \datacloud is very simple: the parameters are
    grouped by collection of properties in SQL tables, such as diameter and 
    albedo (as they are seldom derived independently), mass, thermal inertia, 
    taxonomy, astrometry \citep[the MPCAT-OBS database,][]{mpcat-obs},
    and so on.
    There are a few exceptions to this general scheme.
    The osculating elements of asteroids from the
    Minor Planet Center \citep[MPC,][]{1980CeMec..22...63M}
    and the Lowell observatory \citep{1994IAUS..160..477B}, 
    as well as those of comets from the IMCCE
    \citep{1996IAUS..172..357R}, are stored in separated tables.
    The \Autoref{app:collection} provides the list of collections composing
    the \datacloud ecosystem.

\begin{figure}[t]
      \centering
      \input{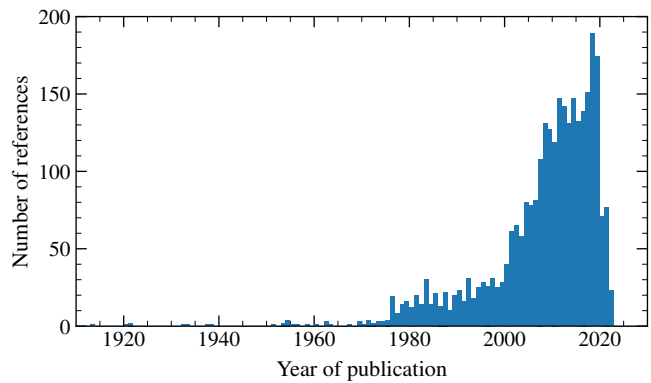}
      \caption{Years of publication of bibliographic references
            used to populate {\ssodnet}.{\datacloud}.}
      \label{fig:datacloud:year}
    \end{figure}

    Each entry of tables corresponds to a single determination of a parameter
    for a given target. Parameters are stored with their uncertainties, the 
    method used to obtain them (see \Autoref{app:method}), a selection flag 
    (used to discriminate among estimates; see \Autoref{sec:ssocard}),
     and the bibliographic reference of the source of data.
    A given SSO, or bibliographic reference, may
    be repeated multiple times: some studies include many objects and
    the same SSO may have been analyzed in multiple studies.
    \Autoref{fig:datacloud:year} shows the distribution over
    time of the publications (currently \numb{\datacloudrefdatasets})
    used to build the \datacloud database.
    For convenience, a file compiling all the bibliographic
    references in \texttt{bibtex} format is 
    available\footnote{\url{https://ssp.imcce.fr/data/ssodnet.bib}}.

    A key aspect of the collections is the unique identifier assigned to each SSO,
    built upon their name and used to identify them across tables. 
    At every update of the database, the name of each SSO (as published by authors)
    is tested with \ssodnet.\quaero and updated upon ingestion. 
    Hence, all properties are linked together using the most 
    up-to-date designation. 

    For each parameter, we started the compilation from scratch, individually adding 
    each bibliographic reference. The only exceptions to this are the masses and the
    spins. For both, we first input a previous compilation of data, taken from
    \citet{2012PSS...73...98C}
    and
    \citet{2021pdss.data...10W}
    respectively.

\section{Selection of the best estimates: \ssocard\label{sec:ssocard}}

  The {\ssodnet}.{\ssocard} provides a practical solution to the question 
  of finding the best estimate for a wide range of parameters of SSOs. From 
  the \datacloud, it builds the resume of each SSO, named \ssocard.
  These \ssocard are small files that can be easily downloaded 
  and read upon user request. The present first release of
  {\ssodnet}.{\ssocard}
  proposes \ssocard for asteroids and dwarf planets only.
  We plan to offer \ssocard for other types of SSOs (comets, satellites)
  in subsequent releases (see \Autoref{sec:future}).

  \subsection{Context}

    Among the hundreds of articles compiled in the \datacloud,
    a significant fraction report the same parameter for a given SSO.
    A question then arises about the most optimal way to choose a value. A simple statistical averaging cannot address the question: 
    some methods are intrinsically more precise than others and some are direct measurements, while others are model-dependent. 
    Moreover, uncertainties associated with values often do not account for 
    possible biases, namely, for external errors. This implies that the choice of the 
    best value cannot entirely rely on criteria that are based on the repeatability of the 
    measurements. 

    The structure and format of data must also be addressed.
    The usual table format (i.e., rows and columns) is not very well adapted to 
    these purposes. Some SSOs have estimates across a wide variety of parameters 
    (osculating elements, proper elements, diameter, mass, density, colors across 
    many filters, taxonomy, and so forth), while others have a few parameters 
    only (e.g., osculating elements). Structuring the data in a flat 2D 
    table implies that a vast majority of cells will be empty. With the 
    current data in \ssodnet, the filling factor of such a table would only be 
    \numb{\bftfillingfactorapprox}  (see \Autoref{sec:bft}).

    Furthermore, the association of data with meta-data (i.e., method, 
    bibliographic reference, and units) is also an issue with regard to the table format.
    Considering that a human-readable bibliographic reference is composed of
    at least four fields (title, authors, year, bibcode), the number of 
    columns will increase by a factor of four for each group of properties. 
    In the current ecosystem of \ssodnet.\datacloud, composed of 
    \numb{\datacloudnbcollections} collections exposing \numb{\datacloudnbfields} 
    fields, it would imply a final table composed of \numb{\datacloudnbmaxcolumns} 
    columns.

    Considering all these elements, we chose to structure the parameters
    in a key-value data format allowing for nested objects and arrays. We chose 
    the open standard file format JSON \citep{STD90_RFC8259}. A XML-based 
    format such as VOTable\footnote{\url{https://ivoa.net/documents/VOTable/}}
    could have been suitable to include metadata, 
    but given it is rather verbose in nature, it would significantly increase the volume 
    of data to exchange.

  \subsection{Method}

    The best estimate for each SSO property depends mainly on the method 
    used to measure it: for example, a direct measurement
    from an in situ space mission 
    can be considered to be more valuable than an indirect 
    determination based on telescopic observations acquired from the Earth.
    Similarly, a modern measurement is often more accurate than an earlier 
    measurement owing to technological advances.
    On the other hand, an old value remains useful because it increases the 
    temporal validity of the measurement and can be unique.
    Finally, the accuracy (closeness to the true value) and precision 
    (repeatability of the value) of measurements must be considered 
    to choose a particular value among a data set or to compute a statistical
    average.

\begin{figure}[t]
      \includegraphics[width=0.5\textwidth]{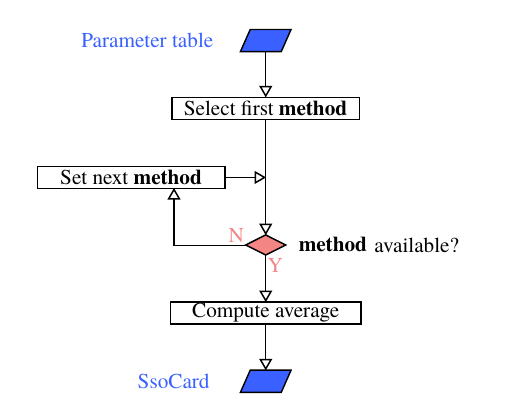}
      \caption{General workflow used to compute the best estimate of
        each parameter.}
      \label{fig:selparam}
    \end{figure}

    For each set of properties, we defined a decision tree that is schematized in
    \Autoref{fig:selparam}. The methods are ordered in a preferential order.
    Among the ordered methods, the first available is chosen, and the
    weighted average $\mu$ is computed from $N$ multiple estimates, $x_i$,
    by the least-squares estimator:
    \begin{equation} 
      \mu = \frac{\sum_{i=1}^{N} w_i\, x_i}{\sum_{i=1}^{N} w_i}
      \label{eq:average}
    ,\end{equation}
    where $w_i = 1 / \sigma_i^2$ and $\sigma_{i} = (\sigma_{+,i} - \sigma_{-,i} )/2$ 
    is the arithmetic mean of the upper and lower uncertainties 
    $\sigma_{+,i}$ and $\sigma_{-,i}$.
    
    Similarly, the upper and lower uncertainties on $\mu$ are computed as:
    \begin{equation}
       \sigma_{\pm} = \frac{\sum_{i=1}^{N} w_i\, \sigma_{\pm,i}}{\sum_{i=1}^{N} w_i}, 
       \ \textrm{with}\  
       w_i = 1 / \sigma_{\pm,i}^2
    .\end{equation}

    When the uncertainty of a value is unknown, we set it to 100\% of 
    the value to weight the mean. At this stage, the $N$ estimates used to compute the average may be less
    than the total number of estimates available. Every single entry
    in \ssodnet.\datacloud has a selection flag (see \Autoref{sec:datacloud}).
    Only three values are possible for this flag: -1, 0 (default), and 1.
    Any estimate with a selection flag of -1 is discarded from the computation
    of the best estimate. If an estimate is flagged with 1, it is considered
    to be the best estimate (we refrain from using it). The overwhelming
    majority of entries in \datacloud have a selection flag of 0.

    We describe below how the preferential order is defined for
    each parameter and we provide the exhaustive order in 
    \Autoref{app:order}.
    Exceptions to this scheme of averaging
    include family membership, albedo, taxonomy, 
    and the orbital elements of SSOs. 

  \subsubsection{Osculating elements\label{ssec:osc}}

    We store in \ssodnet.\datacloud the complete catalogs of
    osculating elements of asteroids and dwarf planets proposed 
    by the MPC \citep[\mpcorb,][]{1980CeMec..22...63M} and the
    Lowell Observatory \citep[\astorb,][]{1994IAUS..160..477B}.
    Osculating elements are a consistent ensemble for each SSO.
    Thus, we do not select them individually, but as a group.
    As the primary source, we chose the \astorb catalog for the \ssocard,
    completed with elements from \mpcorb for SSOs that are not listed in \astorb.

    For each SSO, we used its osculating elements
    (semi-major axis, $a$, inclination, $i$, and eccentricity, $e$) to 
    compute its Tisserand parameter
    \citep{1889BuAsI...6..241T}  
    with Jupiter ($T_J$) and report
    it in the \ssocard:
    \begin{equation}
      T_J = \frac{a_J}{a} + 2\cos{i}\sqrt{\frac{a}{a_J}(1-e^2) }
    ,\end{equation}
    taking $a_J$ = 5.203\,363\,01 au (mean J2000 orbital element).

  \subsubsection{Proper elements}

    Until recently, the only source of proper elements
    was the Asteroid-Dynamical Site\footnote{\url{https://newton.spacedys.com/astdys/}}
    \citep[AstDyS,][]{2003A&A...403.1165K, 2012IAUJD...7P..18K}.
    The computations used either the analytical or numerical methods by 
    \citet{1990CeMDA..49..347M, 1994Icar..107..219M}, 
    \citet{2000CeMDA..78...17K}, and \citet{2002aste.book..603K}.
    More recently, \citet{2019MNRAS.484.3755V} introduced the empirical 
    approach.

    The most recent and largest update on asteroid proper elements is provided by the
    Asteroid Families Portal\footnote{\url{http://asteroids.matf.bg.ac.rs/fam/}}
    \citep{2019EPSC...13.1671N}.
    It thus prevails over the others and we included it in \ssodnet.\datacloud
    to report proper elements of SSOs in \ssocard.
    As Jupiter Trojans and KBOs are not reported in this catalog, we complemented
    it with the proper elements for these populations from AstDyS.

  \subsubsection{Families}

    The existence of asteroid families has been recognized over a century
    ago \citep{1918AJ.....31..185H}. Many authors have been working on the 
    subject over the last decades, using mainly the
    Hierarchical Clustering Method \citep[HCM,][]{1990AJ....100.2030Z}. A new
    method has recently emerged, called 
    V-shape \citep{2017Icar..282..290B}.

    As families are groups of SSOs, the selection is family-based in
    contrast with other parameters that are SSO-based.
    We set as a reference the most-recent large-scale
    study \citep[presently,][]{2019MNRAS.484.3755V}.
    All the families listed in the reference are considered valid and 
    SSOs belonging to these families have a \family item in their
    \ssocard describing their membership.

    We then complete these families with those reported in the other studies
    listed in \ssodnet.\datacloud. We distinguish two cases.
    For articles studying families in general \citep[e.g.,][]{2014Icar..239...46M},
    we add the families not reported in the reference data set. 
    A complexity arises from the fact that different authors may label the same
    family under different names \citep[such as Minerva and Gefion being two names
    pointing at the same family,][]{2014Icar..239...46M, 2015PDSS..234.....N}.
    We thus compute the fraction of common members between reported families.
    Whenever the overlap is smaller than \numb{10\%}, the families are considered 
    different. Alternatively, if one family is significantly smaller than the 
    other (at most \numb{20\%} in number of members), we include it to the list of 
    families as it is likely a sub-family of the larger one.

    For articles focusing on a single family \citep[e.g.,][]{2019MNRAS.482.2612T},
    we consider that they supersede the reference data set. If the family they 
    describe is present in the reference data set, we replace the family membership 
    of all SSOs in the family. If not, we simply add the new family 
    \citep[e.g.,][]{2019A&A...624A..69D}. We illustrate the dynamical families of 
    in the asteroid belt available in \ssodnet in \Autoref{fig:aei_all}.

  \subsubsection{Pairs}

    Pairs of asteroids are objects
    on highly similar heliocentric orbits, first discovered by
    \citet{2008AJ....136..280V}.
    They are similar to dynamical families with only two members and 
    are thought to be formed by rotational fission
    \citep{2007Icar..189..370S, 2010Natur.466.1085P}.
    They are identified from the distance $d$ between
    their orbits (in m.s$^{-1}$):
    \begin{equation}
      \begin{aligned}
        \left(\frac{d}{na}\right)^2 = & \;
           k_a \left( \frac{\Delta a}{a}\right)^2
         + k_e \left( \Delta e\right)^2
         + k_i \left( \Delta \sin i\right)^2 \\
       & + k_\Omega \left( \Delta \Omega\right)^2
         + k_\varpi \left( \Delta \varpi\right)^2 
      \end{aligned}
    ,\end{equation}
    with $\Delta a$, $\Delta e$, $\Delta \sin i$, $\Delta \Omega$, and $\Delta \varpi$
    as the difference in semi-major axis, eccentricity, sine of
    inclination, longitude of the
    ascending node, and argument of perihelion, respectively;
    $n$ and $a$ are the mean motion and semi-major axis of either component; 
    and the numerical constants are
    $k_a = 5/4$,
    $k_e = k_i = 2$, and
    $k_\Omega = k_\varpi = 10^{-4}$ 
    \citep{2019Icar..333..429P}. 
    Backward integration has confirmed many of these pairs, with 
    recent epochs in the past during which the two components were
    within their Hill sphere \citep[see][for instance]{2016A&A...595A..20Z}.
    These epochs are considered the ages of the pairs,
    the time at which the two components became gravitationnally unbound.

    We consider all the pairs listed in the different sources 
    compiled in the \datacloud. However, for the determination of the age,
    for the \ssocard \ we select the most recent determination over
    older studies.

  \subsubsection{Diameter\label{ssec:diam}}

    There are a number different methods available to estimate the diameter of a SSO.
    As a general scheme, we favor estimates obtained by a space mission
    \citep[either via flyby or rendez-vous, such as][]{1992Sci...257.1647B}
    over all the others.
    Diameter estimates based on full 3D shape modeling (including direct
    measurement such as radar echoes, disk-resolved imaging, or 
    stellar occultation) are then considered the most reliable
    \citep[e.g.,][]{1994Sci...263..940H,2010Icar..205..460C,
      2015A&A...576A...8V,2018MNRAS.473.5050B}.

    The next category of methods are 
    convex shape models \citep[generally obtained with the
    light-curve inversion method,][]{2001Icar..153...24K}
    scaled a posteriori using another measurement
    \citep[stellar occultation or mid-infrared 
    flux,][]{2011Icar..214..652D, 1996A&A...310.1011L}
    or tri-axial ellipsoid
    \citep[e.g.,][]{1989Icar...78..323D,2014Icar..236...28D}.
    These are followed by direct measurements limited to a single
    geometry, such as direct imaging \citep{2006Icar..185...39M},
    stellar occultations \citep{1979RMxAA...4..205D},
    interferometry \citep{2009ApJ...694.1228D},
    and broadening of the instrument point-spread function
    \citep{2004AJ....127.2413B}.

    Then come the estimates from the analysis of mid-infrared fluxes with
    spherical models:
    STM \citep{1986Icar...68..239L},
    FRM \citep{1989aste.conf..128L},
    NEATM \citep{1999Icar..142..464H},
    and NESTM \citep{2009MNRAS.400..204W}.
    The last ones chosen are the diameter estimates based on the absolute magnitude,
    $H,$ and the albedo, $p_V,$ (Section~\ref{ssec:albedo})
    when the latter has been derived from the
    polarimetric phase curve of the SSO \citep[e.g.,][]{2007Icar..188..266D}.
    We present the complete list of methods and their order for computing
    the best diameter estimate in \Autoref{tab:rank:diam}.

  \subsubsection{Albedo\label{ssec:albedo}}

    In most cases, the albedo is derived by combining a diameter estimate ($D$)
    with the absolute magnitude, $H,$ at visible wavelengths (more specifically
    in the Johnson V band, hence, the $p_V$ notation), using the canonical
    equation \citep{1989-AsteroidsII-Bowell}:
    \begin{equation}
      p_V = \left(\frac{1329}{D}\right)^2 10^{-0.4 H}\label{eq:albedo}
    .\end{equation}

    An albedo determination is thus closely linked with a diameter estimate and
    this is why both quantities are reported in a single table in
    \ssodnet.\datacloud. Because the absolute magnitude is constantly 
    refined with the new photometry associated with the astrometry reported
    to the MPC, we compute $p_V$ using the latest available absolute 
    magnitude, $H,$ and the best estimate of the diameter (Section~\ref{ssec:diam})
    using \Autoref{eq:albedo}. The uncertainties are computed as:
    \begin{equation}
      \sigma_{\pm,p_V} = p_V \sqrt{ 4 \left( \frac{\sigma_{\mp,D}}{D} \right)^2 +  \left( 0.4 \ln(10) \sigma_{\mp,H} \right)^2 }
    .\end{equation}

\begin{figure*}[t]
      \centering
      \input{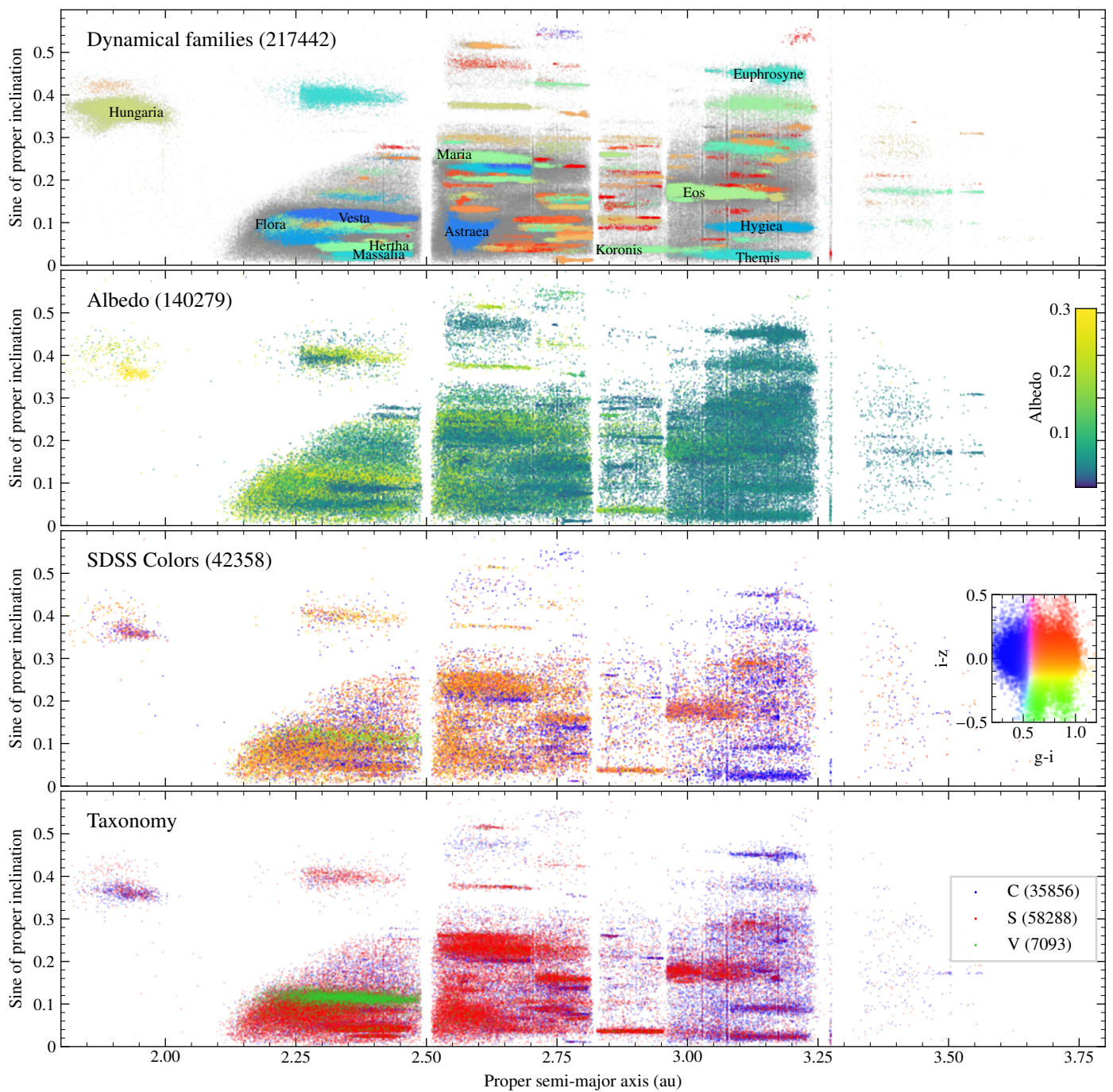}
      \caption{Distribution of families (first panel),
        albedo (second),
        colors (third, using a color-scheme similar to \citet{2008Icar..198..138P}
          based on a code by \citet{astroML}),
        and taxonomy (fourth) against proper elements (semi-major axis and sine of inclination).
        The number of plotted objects is reported in each panel.
      \label{fig:aei_all}}
    \end{figure*}

    Uncertainty on H is seldom provided, and we use a default  value
    of \numb{0.3}.
    The only exceptions to this approach are albedo estimated by
    space missions or, alternatively, from polarimetric phase
    curves (see \Autoref{tab:rank:albedo}), which are not recomputed.
    We present the albedo against proper orbital elements in 
    \Autoref{fig:aei_all}.
 
  \subsubsection{Masses}

    The determination of the mass of an SSO relies on measuring the
    effect of its
    gravitational attraction on another celestial body: either
    a spacecraft or another(s) SSO(s).
    The only exception to this is the mass determination from the 
    detection of Yarkovsky drift \citep{2003Sci...302.1739C}.

    The precision that can be achieved is strongly dependent on the 
    type of interaction, whether that is with: a spacecraft, a satellite in orbit, or
    long-distance encounters \citep{2012PSS...73...98C, 2015-AsteroidsIV-Scheeres}.
    We thus favor mass estimates achieved by radio science experiments
    during spacecraft encounters \citep{1997Sci...278.2106Y, 2011Sci...334..491P}.
    Secondary estimates come from masses determined in binary systems by studying the orbits
    of their moons \citep{
      1999Natur.401..565M, 
      2000Icar..146..190P, 
      2006Sci...314.1276O, 
      2012A&A...543A..68V, 
      2018Icar..309..134P}.
    
    Masses determined from SSO-to-SSO long-distance interactions: close encounters
    \citep{1989Icar...80..326S,2020A&A...633A..46S} and
    ephemerides \citep{2008CeMDA.100...27B, 2008A&A...477..315F} follow.
    Finally, for an SSO with a detected Yarkovsky drift
    \citep{2015-AsteroidsIV-Vokrouhlicky}, it is possible to determine 
    its mass on the basis of a number of other parameters
    \citep[diameter, albedo, obliquity, thermal inertia, etc., as per][]{2014Icar..235....5C}.
    We present the complete ordered list of methods for computing
    the best mass estimate in    
    \Autoref{tab:rank:mass}.

  \subsubsection{Density}

\begin{figure}[t]
      \centering
      \input{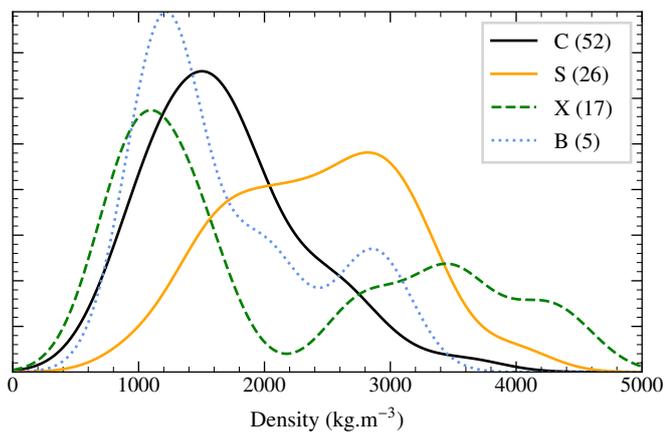}
      \caption{Kernel density estimate (KDE) of the density of the
        C, S, X, and B complexes. The bimodal distribution of 
        X-types highlights the P and M sub-classes (average $p_V$ of
        \numb{0.044} and \numb{0.129}, below and above 2000 kg.m$^{-3}$,
        respectively).
        Similarly, Pallas is the sole contributor to high-density B-types.
      \label{fig:density}}
    \end{figure}

    For each SSO with both a mass $M$ and a diameter $D$ estimates, we
    compute its density $\rho$ (kg.m$^{-3}$) and associated uncertainties as follows:
    \begin{eqnarray}
      \rho &=& M \left/ \frac{\pi}{6} D^3 \right. \\ 
      \sigma_{\pm,\rho} &=& \rho
      \sqrt{ 9 \left(
         \frac{\sigma_{\mp,D}}{D}
         \right)^2 +
         \left(
         \frac{\sigma_{\pm,M}}{M}
         \right)^2 
      }
    \end{eqnarray}
    In some cases, the density can be determined without knowledge of
    either the mass or the volume. This is often the case of small
    binary asteroid systems studied by optical light curves
    \citep{2009Icar..200..531S, 2015Icar..248..516C}. A few binary
    systems imaged by radar are also included in this case
    \citep{2014DPS....4621315F}.
    Last, the density can be derived from a detected Yarkovsky
    orbital drift \citep{2014A&A...568A..43R}.
    We did not set preference of a method over another and we chose to average these
    estimates together. The distribution of density for a few
    selected taxonomic classes is presented in \Autoref{fig:density}.
 
  \subsubsection{Spin solutions}

    In most cases, the only available information on the spin of
    an SSO is its rotation period (often reported as synodic period).
    In some cases, however, the orientation of the spin axis has
    been determined, and we report its coordinates
    both in ECJ2000 \citep[as reference time, longitude, and latitude;
      see][]{2001Icar..153...24K, 2010A&A...513A..46D}
    and in EQJ2000
    \citep[as right ascension, declination, and the position of the
      prime meridian $W_0$ and $\dot{W}$; see][]{2018CeMDA.130...22A}.
    
\begin{figure}[t]
      \centering
      \input{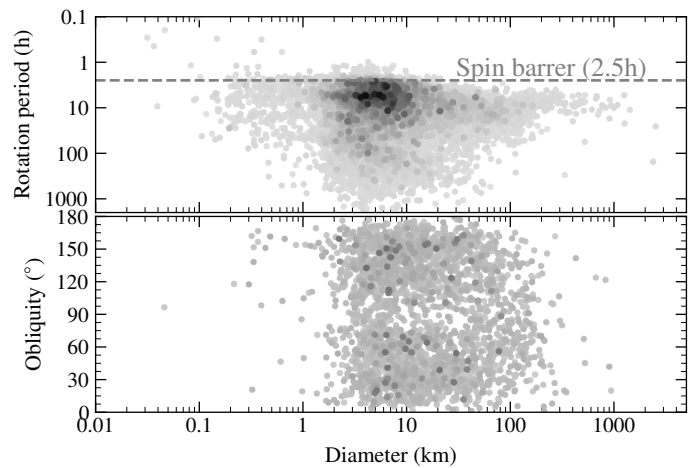}
      \caption{Rotation period (top) and obliquity
        (bottom) vs diameter for \numb{17,201} and
        \numb{2596} SSOs, respectively.
        Darker shades of grey indicate higher density of points.
      \label{fig:spin}}
    \end{figure}

    Spin-vector coordinates determined with the light-curve inversion
    method \citep{2001Icar..153...24K} are often degenerated with a 
    mirror solution separated by 180\degr\ in ecliptic longitude.
    We use the selection flag (\Autoref{sec:datacloud}) to remove this 
    ambiguity whenever one of the two spin solutions has been rejected
    a posteriori
    \citep[from comparison with stellar occultation or disk-resolved
      imaging for instance,][]{2006Icar..185...39M, 2011Icar..214..652D}.
    For each SSO with spin-vector coordinates, we computed its obliquity
    using these coordinates and its osculating elements (Section~\ref{ssec:osc}).
    We present the distribution of rotation period and obliquity against
    diameter in \Autoref{fig:spin}.
 
    Here, again, solutions obtained by spacecraft encounters are favored over
    any others. They are followed by spin solutions obtained by 3D shape
    modeling techniques that include direct disk-resolved measurements
    \citep[stellar occultations,
      disk-resolved images, etc., e.g.,][]{2015MNRAS.448.3382T, 2018A&A...618A.154V,
      2018Icar..311..197S, 2019A&A...623A.132C}.
    These are followed by 3D shape models that are later scaled using complementary
    observations \citep[mid-infrared fluxes, stellar occultations,
      disk-resolved images, etc.,][]{2013Icar..226.1045H, 2011Icar..214..652D}.
    Spin solutions associated with convex shape models, generally 
    with a mirrored spin solution, were then chosen
    \citep{2013A&A...559A.134H, 2018A&A...610A...7M}, followed
    by solutions obtained from tri-axial ellipsoids
    \citep{1989Icar...78..323D, 2013Icar..225..794M}.
    Finally, there are the periods determined from light curves, 
    with or without constraints on the spin coordinates
    \citep{1978A&AS...34..203L, 2020AJ....160...73Y}.
    We refer to \Autoref{tab:rank:spin} for a full listing of the 
    order of preference. 

    \begin{figure}[t]
      \centering
      \input{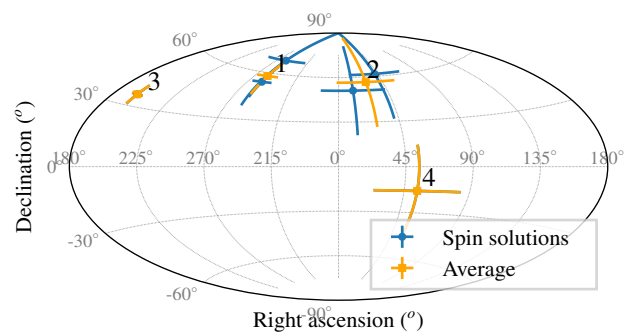}
      \caption{Example of clustering of spin coordinates for
              \nuna{20}{Massalia}. Six solutions (blue) are possible
              {\citep{2002Icar..159..369K, 
              2016A&A...586A.108H,
              2019A&A...631A..67C}}.
              The four separated spin coordinates (orange) of the four
              clusters are reported in the \ssocard.
      \label{fig:spincoords}}
    \end{figure}

    The average spin coordinates are computed using \Autoref{eq:average}.
    However, as several ambiguous spin solutions may co-exist for a given
    SSO, we identify which estimates correspond to which spin solution using
    K-Means clustering \citep{KMeans}, as provided by the
    \texttt{scikit-learn}\footnote{\url{https://scikit-learn.org}}
    python package \citep{scikit-learn}.
    We consider that up to four distinct spin solutions can be present, such
    as for \nuna{20}{Massalia} (\Autoref{fig:spincoords}).
    Spin coordinates must be within \numb{30}\degr~of
    the average to be include in a cluster.
    We set default uncertainties of \numb{30}\degr~on
    spin coordinates whenever they have not been specified by the respective authors.
We used a similar approach, based on K-Means clustering, for the rotation periods.
    In this case, the threshold to belong to a solution was set to \numb{0.2}\,h.
    The default uncertainty was set to \numb{1}\,h.

  \subsubsection{Colors}

    Stricto sensu, the colors of SSOs are observable and not derived properties.
    Nevertheless, we compiled the colors of SSOs in \ssodnet.\datacloud, with the same
    rationale as for derived properties: many colors are available but spread
    over many studies \citep[e.g.,][]{2003Icar..163..363D,2010A&A...511A..72S,
      2018AN....339..198D} and they are    usually not in machine-readable format. Furthermore, colors
    can be used for taxonomic determination
    \citep{2010A&A...510A..43C, 2013Icar..226..723D}.

    Several ancillary information for contextualization are recorded
    (\Autoref{tab:struct:colors}), such as the observing time, the source 
    of measurement (plain English  description and IAU Observatory 
    code\footnote{\url{https://minorplanetcenter.net/iau/lists/ObsCodesF.html}}
    if available).
    The filters used to compute the colors are identified with the unique
    identifier of the SVO Filter Service\footnote{\url{http://svo2.cab.inta-csic.es/theory/fps/}}
    providing transmission curves and zero points
    \citep{2012ivoa.rept.1015R,2020sea..confE.182R}. Similarly, we record
    in which system the photometry is reported (Vega, AB, or ST).

    The selection of best estimates is based on 
    the time difference, $\Delta_t$, between the observation
    of the two filters and how the color was computed.
    We favor (\Autoref{tab:rank:color}) colors computed 
    as a difference of absolute magnitudes from phase functions 
    in each filter \citep{2021Icar..35414094M, AlvarezCandal2021}. 
    In that case, we report the most-recent published value.
    Then we follow up with the colors computed as a difference of apparent magnitudes but corrected for light curve
    variations \citep{2016AJ....151...98M, 2019ApJS..242...15E}.
    Last, we have  the simple difference of apparent magnitudes
    \citep{2018A&A...617A..12P, 2021A&A...652A..59S}.
    Whenever several estimates of the same color with the two latter methods are reported,
    we computed their average as in \Autoref{eq:average}, with the following weight to account
    for time difference: 
    $w_i = 1 / \sigma_i^2 + 1/ \Delta_t^2$.
    Whenever the information on $\Delta_t$ is missing, we set it to \numb{1\,h}.

    Last but not least, filter transmissions are different in each facility.
    For a given color (e.g., $g$-$i$), the values from different observatories may differ 
    \citep[e.g., between the Sloan Digital Sky Survey and SkyMapper, see Fig.~9 in][]{2022A&A...658A.109S}.
    We did not merge colors obtained with different filter sets, for instance,
    \texttt{SLOAN/SDSS (g$\,$-$\,$i)} vs. \texttt{SkyMapper/SkyMapper (g$\,$-$\,$i)},
    but instead we report the most precise results. An example of these colors is shown 
    in \Autoref{fig:aei_all}.

   \subsubsection{Phase function}

    Phase functions describe the evolution of brightness with 
    the phase angle (once it is corrected with respect to 
    the Sun-target and target-observer distances).
    The absolute magnitude reported together with osculating elements
    (Section~\ref{ssec:osc}) is computed using the historical two-parameter
    HG phase function \citep{1989-AsteroidsII-Bowell}, where G is 
    generally assumed to be 0.15.
    This function has been shown to deviate from observed photometry at
    low and high phase angle, and a three-parameter HG$_1$G$_2$ function
    has been proposed \citep{2010Icar..209..542M}. We collect 
    these parameters in the \datacloud and report them in \ssocard. 
    Because phase functions are wavelength-dependent
    \citep{2012Icar..220...36S, 2021Icar..35414094M}, we associate
    these parameters with the filter in which they were derived,
    again using the unique
    identifier of the SVO Filter Service
    \citep{2012ivoa.rept.1015R,2020sea..confE.182R}. 

    A parameterized version of the phase function has
    been proposed for low-accuracy data \citep[with two parameters, 
    HG$_{12}$, later refined as HG$_{12}^\star$,][]{2016&PSS..123..117P}.
    However, we stick to HG$_1$G$_2$ parameters only,
    as they have been shown to
    convey taxonomic and albedo information
    \citep{2016P&SS..123..101S,2021Icar..35414094M}.

  \subsubsection{Taxonomy}

    Taxonomy is often used as a proxy for composition
    in statistical studies of populations \citep{2008Icar..198..138P, 
      2014Natur.505..629D, 2019Icar..324...41B, 2021ApJ...916L...6H}.
    The complexity of compiling taxonomic classes is manifold. First,
    several taxonomies (as classification schemes) have been developed
    and used by the community, such as
    \citet{1989aste.conf.1151T},
    \citet{1989aste.conf.1139T}, 
    \citet{2002Icar..158..146B}, 
    \citet{2009Icar..202..160D},
    and 
    \citet{2022arXiv220311229M}.
    Second, there is a great diversity in the potential combinations
    of these schemes with
    observing techniques 
    \citep[multi-filter photometry and 
      spectroscopy,][]{1995Icar..115....1X, 2016Icar..268..340C}
    and wavelength    \citep[visible only, near-infrared only, and both,][]{2010A&A...510A..43C, 2018A&A...617A..12P, 2014A&A...568L...7M}.

\begin{figure*}[ht]
      \includegraphics[width=\textwidth]{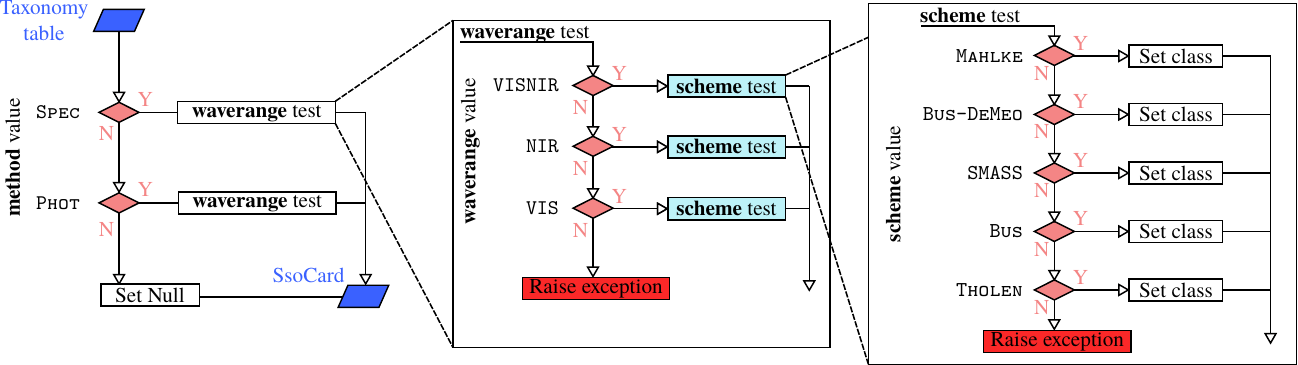}
      \caption{Decision tree for taxonomic classes. Classes based on 
         spectroscopy are favored upon those based on broad-band
         photometry only. Similarly data sets covering the full
         VISNIR wavelength range are favored over NIR only, 
         itself preferred to VIS only. Finally, for classification 
         based on similar data sets, the Mahlke scheme is preferred 
         over Bus-DeMeo, Bus, SMASS, and Tholen schemes.
         \label{fig:seltaxo}
      }
    \end{figure*}

    We present in \Autoref{fig:seltaxo} the decision tree we applied to 
    select the most relevant taxonomy for a given SSO.
    As a general rule, results from spectroscopy are favored over
    results from multi-filter photometry. Within each observing 
    technique, the results using both visible and near-infrared are
    favored, then those based on infrared only, and then finally those based on
    the visible only. Once the observing technique and wavelength range 
    is selected, there may be several taxonomic schemes available, and
    we chose the Mahlke, Bus-DeMeo, SMASS, Bus, and Tholen taxonomies, respectively.

    In an attempt to homogenize all the classes that have been reported
    for a given object, we also group similar classes under the term
    ``complex,'' following the associations listed in \Autoref{tab:taxo}.
    We give an example of the orbital distribution of these
    complexes in \Autoref{fig:aei_all}.

  \subsubsection{Thermal properties}

    Mid-infrared fluxes are often used to determine the diameter of an
    SSO (Section~\ref{ssec:diam}), from simple thermal models such as NEATM \citep{1999Icar..142..464H}.
    More complex thermal models \citep[referred to as thermophysical
      models, TPM,][]{1996A&A...310.1011L}
    can also be used, but require additional information on the object
    such as spin, 3D shape, and so on. One parameter used in TPM is the 
    thermal inertia (in J.s$^{-1/2}$.K$^{-1}$.m$^{-2}$) controlling the
    resistance of the surface to changes of temperature.

\begin{figure}[t]
      \centering
      \input{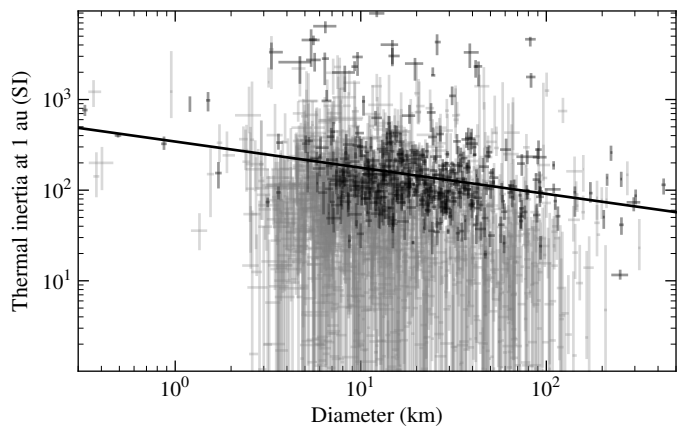}
      \caption{All \numb{1681} SSOs with a thermal inertia above 1\,SI (gray),
               the \numb{419} with a S/N above 3 (black),
               and a linear regression on the latter of equation
               $\log(\Gamma_0) = 2.5 - 0.29\log(D)$, a result that is similar to the
               recent work of
               \citet{2022PSJ.....3...56H}.
      \label{fig:TI}}
    \end{figure}

    The thermal inertia determination from spacecrafts are favored
    \citep{2014GeoRL..41.1438C}, followed by
    those determined from TPM using a priori knowledge on the spin and shape
    \citep{2013Icar..226..419M, 2012P&SS...66..192O},
    and, finally, the TPM applied to spheres \citep{2013A&A...558A..97M},
    as listed in \Autoref{tab:rank:thermal}.
    Thermal inertia ($\Gamma$) is a function of heliocentric distance
    \citep{1999Icar..141..179V,2018MNRAS.477.1782R}. We thus report
    the thermal inertia at 1\,au ($\Gamma_0$) from the Sun in the \ssocard, using
    the following relation:
    \begin{equation}
      \Gamma = \Gamma_0 r_H^\alpha
    ,\end{equation}
    where $r_H$ is the heliocentric distance at the time of the observations and we take $\alpha$\,=\,$-3/4$ following \citet{2015-AsteroidsIV-Delbo}.
    We present the distribution of thermal inertia
    against diameter 
    in \Autoref{fig:TI}.

  \subsubsection{Yarkovsky drift}

    While the orbital drift due to the delayed thermal emission by
    asteroid surface is extremely small
    \citep[of the order of $10^{-4}$ au/Myr,][]{2015-AsteroidsIV-Vokrouhlicky},
    it was detected for the first time almost two decades ago
    \citep{2003Sci...302.1739C}.
    We favor detections that include both optical and radar observations
    \citep[][for instance]{2014CeMDA.119..301F}
    over those using optical only \citep[e.g.,][]{2018A&A...617A..61D}.
    Finally, we consider those (\Autoref{tab:rank:yarko}) estimated based on the age of dynamical families
    \citep{2017MNRAS.469.4400C}.

    Some authors have reported the semi-major axis drift, $\dot{a}$
    \citep[][]{2012AJ....144...60N}, while
    others have given the transverse acceleration, $A_2$
    \citep[][]{2019Icar..321..564G},
    as in the case of cometary dynamical models \citep{1973AJ.....78..211M}.
    We report both parameters in the \ssocard,
    using the following equation from \citet{2013Icar..224....1F}
    to convert between quantities:
    \begin{equation}
      \dot{a} = \frac{ A_2 }{ a^2 (1-e^2) \pi n}
                \int_0^{2\pi} (1 + e \cos f) df
    ,\end{equation}
    with $a$ as the semi-major axis, $e$ as the eccentricity,
    and $n$ as the mean motion (Section~\ref{ssec:osc}).

\section{Summary for all SSOs: \bft\label{sec:bft}}

  The \ssocard service described in previous section provides
convenient access to the best estimates of many parameters, but limited
  to a single SSO. The last service composing \ssodnet thus provides
  a broad and flat table (\bft) that compiles all the parameters of the
  \ssocard for all SSOs.
  This table is very large (over \numb{\datacloudnbfields} fields for 
  \numb{\bftnbrows} SSOs, about \numb{\bftfilesize}\,Gb). Yet, most fields are 
  empty (i.e., there is no estimate of the given parameter for this SSO), 
  resulting in only a \numb{\bftfillingfactor} filling factor.

  We propose the \bft as an enhanced character separated values
  (eCSV\footnote{\url{https://github.com/astropy/astropy-APEs/blob/main/APE6.rst}})
  and an Apache parquet\footnote{\url{https://parquet.apache.org/}}
  files for users interested in the statistical properties of the asteroid 
  population. These files can be downloaded at static urls
  (eCSV\footnote{\url{https://ssp.imcce.fr/data/ssoBFT-latest.ecsv.bz2}},
  parquet\footnote{\url{https://ssp.imcce.fr/data/ssoBFT-latest.parquet}}).
  We also provide this table to the CDS to ensure its fully VO-compliant 
  access.

\section{Accessing the services: \ssodnet \& \rocks\label{sec:access}}

  We offer several access interfaces to the \ssodnet service, described below.

  \subsection{REST interface\label{sec:rest}}

    The \quaero representational state transfer (REST)
    API is a low-level interface dedicated to developers. It is designed 
    to offer an easy-to-use and fast solution to search for planetary objects (\texttt{sso} 
    and \texttt{search} methods) to resolve their designations (\texttt{resolver} method) 
    or to be used as an auto-completion mechanism for names (\texttt{instant search} method)
    into Web forms and applications connected to the Internet.
    In the framework of the Virtual Observatory, no standard protocol nor technical 
    specification is quite capable of designing a fast-search engine. Thus, the core 
    of \ssodnet name resolver does not follow any current VO standard. Nevertheless, the 
    underlying technology and the API we have chosen being intrinsically interoperable, 
    the \quaero service can easily be included in any VO ecosystem.

	\begin{table}[h!]
	  \centering
	  \small
	  \begin{tabular}{lp{0.37\textwidth}}
		\hline 
        End-point: & \url{https://api.ssodnet.imcce.fr/quaero/1/} \\
        Doc:       & \url{https://doc.ssodnet.imcce.fr/quaero.html} \\
		\hline 
	  \end{tabular} 
	\end{table}

  \subsection{Web-service interface\label{sec:ws}}

    We provide a Web-service interface, built upon XML and SOAP technology,
    that allows for a full interaction with \ssodnet through several methods:
    \begin{enumerate*}[label=(\roman*)]
       \item \texttt{resolver}: to identify SSO (high level API),
       \item \texttt{datacloud}: to retrieve all known values of SSO properties,
       \item \texttt{ssocard}: to retrieve the best estimates of SSO properties.
    \end{enumerate*}
    The user can simply post a request to the method end-points to gather 
    corresponding data, using a data transfer program such as \texttt{wget} 
    or \texttt{curl}.
    More advanced users can implement the SOAP Web service to ensure
    an application-to-application communication between \ssodnet and
    a software or a public Web page.

	\begin{table}[h!]
	  \centering
	  \small
	  \begin{tabular}{lp{0.32\textwidth}}
		\hline 
        \ssodnet server:  & \url{https://ssp.imcce.fr/webservices/ssodnet/ssodnet.php} \\
        Public interface: & \url{https://ssp.imcce.fr/webservices/ssodnet/ssodnet.php?wsdl} \\
        Doc:              & \url{https://ssp.imcce.fr/webservices/ssodnet/} \\
		\hline 
	  \end{tabular} 
    \end{table}

  \subsection{Web form interface\label{sec:ws}}

    The easiest way to search for a SSO and to quickly consult its properties 
    may be to use \ssodnet dedicated Web form. The best estimates of the physical 
    and dynamic properties (the \ssocard) are displayed in a comprehensive manner, 
    together with bibliographic references.
    We also provide links to all values (i.e., \datacloud entry for each property
    of the SSO), and to the subset used to compute the best estimates (as
    defined by the decision trees, see \Autoref{sec:ssocard}).

	\begin{table}[h!]
	  \centering
	  \small
	  \begin{tabular}{lp{0.37\textwidth}}
		\hline 
        Web form: & \url{https://ssp.imcce.fr/forms/ssocard/} \\
        Doc:      & \url{https://ssp.imcce.fr/forms/ssocard/doc} \\
		\hline 
	  \end{tabular} 
    \end{table}

  \subsection{Python interface: \rocks\label{sec:rocks}}

    We provide a \texttt{python} interface to \ssodnet named \rocks.
    It offers a programmatic entry point both for data exploration and data
    processing. The interaction with the \ssodnet repositories is asynchronous
    and results are cached on the user-side, providing a responsive user
    experience.

	\begin{table}[h!]
	  \centering
	  \small
	  \begin{tabular}{lp{0.37\textwidth}}
		\hline 
        Sources: & \url{\urlRocks} \\
        Doc:     & \url{\urlRocksDocs} \\
		\hline 
	  \end{tabular} 
    \end{table}

    Data exploration is accessible via the command line interface of \rocks in a
    straightforward, uniform syntax:
    \begin{verbatim}
$ rocks [command|parameter] [asteroid_identifier]
    \end{verbatim}

    Here, the \texttt{parameter} can be any key from the \ssocard or \datacloud
    catalogs, while the \texttt{asteroid\_identifier} is any identifier that can
    be resolved by \quaero. The result of the query is printed in the
    console. Commands such as \texttt{id} and \texttt{info} serve to identify an
    asteroid and to print the asteroid's \ssocard.

    \begin{verbatim}
$ rocks id "1975 XP"
(234) Barbara 
$ rocks taxonomy Barbara
L
$ rocks diameter barbara
46.3 +- 5.0 km
$ rocks albedo 234  
0.187 +- 0.2839 
    \end{verbatim}

    An overview of all compiled literature values is printed when requesting the
    plural of the parameters. This is possible for all parameters which have
    \datacloud entries, such as albedo, mass, taxonomy, etc.

    \begin{verbatim}
$ rocks taxonomies ceres
+-------+--------+-----------+-----------------+
| class | method | scheme    | shortbib        |
+-------+--------+-----------+-----------------+
| G     | Phot   | Tholen    | Tholen+1989     |
| C     | Spec   | Bus       | Bus&Binzel+2002 |
| C     | Spec   | Bus       | Lazzaro+2004    |
| C     | Spec   | Tholen    | Lazzaro+2004    |
| C     | Spec   | Bus-DeMeo | DeMeo+2009      |
| C     | Spec   | Bus       | Fornasier+2014b |
| G     | Spec   | Tholen    | Fornasier+2014b |
| C     | Phot   | Bus-DeMeo | Sergeyev+2022   |
| C     | Spec   | Mahlke    | Mahlke+2022     |
+-------+--------+-----------+-----------------+
    \end{verbatim}

    Data processing is facilitated for \texttt{python} scripts using the \rocks
    package. The main entry point is the \texttt{rocks.Rock} class, where each
    instance reflects a unique asteroid. The asteroid parameters are accessible
    as class attributes via the dot notation, which again leads to an
    intuitive syntax:

    \begin{verbatim}
    >>> import rocks
    >>> vesta = rocks.Rock(4)
    >>> vesta.albedo.value
    0.38
    >>> vesta.albedo.error.min_
    -0.04
    >>> vesta.albedo.error.max_
    0.04
    >>> vesta.albedo.description
    'Geometrical albedo in V band'
    \end{verbatim}

    The asynchronous interaction with the locally cached data and the remote
    \ssodnet repositories allow for a fast analysis process without the use of 
    resource-intensive multiprocessing or multi-threading strategies. To provide an
    estimate of the execution times, we identified all asteroids in the SDSS
    Moving Object Catalog
    DR1\,\footnote{\url{http://faculty.washington.edu/ivezic/sdssmoc/sdssmoc.html}}
    and retrieved their \ssocard. The catalog contains observations
    of \numb{10,585} unique minor bodies, largely referred to by designations that
    are no longer the main identifier of the object. Using a combination of
    \quaero queries and a local asteroid name-number-designation index, \rocks
    identifies all objects within \numb{2.5}\,s. The \texttt{ssoCards} are retrieved
    within \numb{320}\,s from \ssodnet, about 30\,ms per asteroid. \rocks then performs data
    validation and deserialization (i.e.,{} converting the JSON server response
    into a \texttt{python} object) within \numb{120}\,s, that is, about 11\,ms per asteroid.
    A second execution of the analysis script would benefit from the locally
    cached \texttt{ssoCards}, rendering any request to \ssodnet obsolete.

    To install \rocks, we can use the \texttt{python} package index (PyPI) under
    the package name \texttt{space-rocks}.
    The online documentation\footnote{\url{\urlRocksDocs}}
    provides a guide on getting started and tutorials to achieve more advanced data
    processing results. We note that \rocks is actively developed and maintained by 
    the authors of this work.

\section{Future developments\label{sec:future}}

  We foresee several lines of development for the \ssodnet service:
  data compilation and curation, expansion of the set of parameters and 
  types of SSOs, and development of the interface.

  \paragraph{Data compilation:} 
  First and foremost, we will continue to compile data into the \datacloud, aiming
  for completeness with respect to the listed parameters. Indeed, it is the building block 
  of the \ssocard and the \bft, which are automatically generated from the 
  entries in the \datacloud. On the other hand, \quaero has been working
  and been updated weekly for several years, following the growing list of
  SSOs listed by the MPC. Thus, a continuous scientific monitoring of publications
  is required for the service.

  We welcome any feedback, especially on data sources that may
  be missing or erroneous entries. While we conducted multiple checks on the
  data included in \ssodnet, some typographical errors may lurk in 
  the unprecedented size of the data compilation. We will happily include
  sources that are not included in the current release of the service
  and correct entries.

  Furthermore, \ssodnet can be used by any group or researcher to
  publish regularly updated data. A simple file (VOTable, csv, ...), with sufficient metadata
  at a static url can be used as a source, without requiring
  a server or a database with a web service.

  \paragraph{Set of parameters:} The set of parameters currently available
  in \ssodnet is already broad, covering dynamical, surface, and physical
  properties (\Autoref{tab:collections}). There are, however, other parameters
  of interest that will be added to the \datacloud (and, hence, \ssocard
  and \bft), such as
  the source region probabilities for near-Earth objects
  \citep{2017A&A...598A..52G} and their minimal orbital intersection distance 
  with planets \citep[MOID,][]{1993-Marsden_MOID},
  activity for asteroids and Centaurs
  \citep{2006Sci...312..561H, 2009AJ....137.4296J},
  and radar albedos \citep{2014Icar..238...37N}.
  Additional computed parameters can also be added in \ssocard, such as
  surface gravity or escape velocity.

  \paragraph{Types of SSOs:} The present release of \ssodnet focuses on 
  asteroids because they are the prime targets of study of the authors. 
  The service was nevertheless designed to cope with all classes of 
  SSOs: comets, planets, satellites, and interstellar objects.
  For instance, \quaero already deals with the designation of all these 
  categories.

  We thus welcome partnership with everyone willing to contribute to
  build this community database. Beside the collection and curation
  of data, a set of parameters relevant for these celestial
  bodies must be defined (e.g., non-gravitational acceleration for comets, 
  libration amplitude and frequency for satellites),
  together with decision trees to estimate the best parameters.
  \ssodnet has been envisioned as a
  service to the community and any contribution to it will expand its advantages.

  \paragraph{User interface:}
  \ssodnet is mainly a machine-machine service,
  allowing for on-the-fly data retrieval. 
  Both \quaero and \ssocard are designed to cope with constant
  queries. The \datacloud entries for a given SSO can also be dumped
  easily, and the \bft downloaded as a whole.

  We plan to develop more advanced possibilities to query the data, both
  in \datacloud and \bft. Users may be interested by searching
  entries from a given bibliographic reference, rather than
  for a specific SSO for instance.
  Similarly, users may be interested in a subset
  of the \bft only (e.g., some specific parameters only for SSOs fulfilling
  certain conditions). While the latter is possible with TAP on the
  version of the \bft hosted at the CDS, the former requires development
  on the server side of \ssodnet.

\section{Conclusions}

  We present a new Web Service, \ssodnet, which provides a convenient
  solution to the issues of SSO identification and the compilation of 
  properties.
  It consists of a suite of applications,
  each with its own programming interface:
  \quaero for name resolution, 
  \datacloud compiling SSO properties,
  \ssocard providing the set of best estimates for each SSO, and
  \bft compiling the latter for all SSOS.
  These entry points deliver JSON as native outputs. We have released a
  \texttt{python} interface for these services: \rocks,   
  available in the \texttt{python} package index (PyPI).
  \ssodnet is fully operational. The name resolver \quaero is updated
  weekly to follow SSO discoveries. We plan on monthly updates for
  the others applications, following compilation of data from continuous
  monitoring of new publications.
  The future evolution of the service includes an extension of the suite
  of properties and classes of SSOs, along with an advanced query interface
  to retrieve large corpus of data.

\begin{acknowledgements}
  This research has made use of the SVO Filter Profile
  Service supported from the Spanish MINECO through grant AYA2017-84089
  \citep{2012ivoa.rept.1015R, 2020sea..confE.182R}.
  This research has made use of the \texttt{scikit-learn}
  \texttt{python} package \citep{scikit-learn}.
  We did an extensive use of the VO TOPCAT software
  \citep{2005ASPC..347...29T}.
  Thanks to all the developers and maintainers.
  We would like to thank J.~Masiero, F.~E.~DeMeo, and
  F.~Spoto for discussions
  that led to the current decision trees used in \ssodnet.
\end{acknowledgements}

\bibliographystyle{aa} 
\bibliography{main}


\clearpage
\onecolumn

\begin{appendix}

   \renewcommand{\arraystretch}{1.2}

   \section{Collections available in {\ssodnet}.{\datacloud}\label{app:collection}}
   \begin{longtable}{lLrrll}
      \caption{Description of the collections included in {\ssodnet}.{\datacloud}.\label{tab:collections}}\\
      \hline\hline
      Name & Description & $N$ & $N_{\textrm{SSO}}$ & Desc. & Reference \\
      \hline
      \endfirsthead
      \caption{continued.}\\
      \hline\hline
      Name & Description & $N$ & $N_{\textrm{SSO}}$ & Desc. & Reference \\
      \hline
      \endhead
      \hline
      \multicolumn{2}{r}{Total} & \numb{351\,083\,883} & \numb{\bftnbrows} & & \numb{\datacloudrefdatasets} references \\
      \hline
      \multicolumn{6}{p{0.95\columnwidth}}{For each collection we list the number of entries ($N$), number of SSOs ($N_{\textrm{SSO}}$),
                the reference to a table describing its fields (Desc.), and the number of included bibliographic references.}\\
      \endfoot
      \astorb      & Lowell orbits of asteroids     & \numb{\astorbN}         & \numb{\astorbNSSO}         & \ref{tab:struct:astorb} & \citet{1994IAUS..160..477B} \\
      \colors      & Compilation of colors          & \numb{\colorsN}         & \numb{\colorsNSSO}         & \ref{tab:struct:colors} & \numb{\colorsNBibRef} references \\
      \cometpro    & IMCCE orbits of comets         & \numb{\cometproN}       & \numb{\cometproNSSO}       & \ref{tab:struct:cometpro}&\citet{1996IAUS..172..357R} \\
      \density     & Density estimates              & \numb{\densityN}        & \numb{\densityNSSO}        & \ref{tab:struct:density}& \numb{\densityNBibRef} references \\
      \diamalbedo  & Diameter \& albedo estimates   & \numb{\diamalbedoN}     & \numb{\diamalbedoNSSO}     & \ref{tab:struct:diam}   & \numb{\diamalbedoNBibRef} references \\
      \families    & Dynamical families             & \numb{\familiesN}       & \numb{\familiesNSSO}       & \ref{tab:struct:fams}   & \numb{\familiesNBibRef} references \\
      \masses      & Mass estimates                 & \numb{\massesN}         & \numb{\massesNSSO}         & \ref{tab:struct:mass}   & \numb{\massesNBibRef} references \\
      \mpcatobs    & MPC catalog of observations    & \numb{\mpcatobsN}       & \numb{\mpcatobsNSSO}       & \ref{tab:struct:mpcat}  & \citet{mpcat-obs} \\
      \mpcorb      & MPC orbits of asteroids        & \numb{\mpcorbN}         & \numb{\mpcorbNSSO}         & \ref{tab:struct:mpcorb} & \citet{1980CeMec..22...63M} \\
      \pairs       & Asteroid pairs                 & \numb{\pairsN}          & \numb{\pairsNSSO}          & \ref{tab:struct:pairs}  & \numb{\pairsNBibRef} references \\
      \phase       & Parameters of phase functions  & \numb{\phasefunctionN}  & \numb{\phasefunctionNSSO}  & \ref{tab:struct:phase}  & \numb{\phasefunctionNBibRef} references \\
      \proper      & Proper elements of asteroids   & \numb{\properelementsN} & \numb{\properelementsNSSO} & \ref{tab:struct:proper} & \citet{2019EPSC...13.1671N} \\
      \spin        & Spin solutions                 & \numb{\spinN}           & \numb{\spinNSSO}           & \ref{tab:struct:spin}   & \numb{\spinNBibRef} references \\
      \taxonomy    & Taxonomic classes              & \numb{\taxonomyN}       & \numb{\taxonomyNSSO}       & \ref{tab:struct:taxo}   & \numb{\taxonomyNBibRef} references \\
      \thermal     & Thermal inertia estimates      & \numb{\thermalinertiaN} & \numb{\thermalinertiaNSSO} & \ref{tab:struct:thermal}& \numb{\thermalpropertiesNBibRef} references \\
      \yarko       & Yarkovsky drifts               & \numb{\yarkovskyN}      & \numb{\yarkovskyNSSO}      & \ref{tab:struct:yarko}  & \numb{\yarkovskyNBibRef} references \\
   \end{longtable}

   \qquad
 
   \begin{longtable}{rllQ}
      \caption{Description of the fields in the collection \astorb of the \datacloud.\label{tab:struct:astorb}}\\
      \hline\hline
      \# & Field & Type & Description \\
      \hline
      \endfirsthead
      \caption{continued.}\\
      \hline\hline
      \# & Field & Type & Description \\
      \hline
      \endhead
      \hline
      \endfoot
      \hline
      \multicolumn{4}{p{0.95\columnwidth}}{Fields follow the original ASTORB data \citep{1994IAUS..160..477B}, and we 
                refer to the online documentation for further details on each field (\url{https://asteroid.lowell.edu/main/astorb/}).}\\
      \endlastfoot
      1 & num                 & int     & SSO IAU number \\
      2 & name                & varchar & SSO name \\ 
      3 & orbit\_computer     & varchar & Orbit computer \\
      4 & H                   & double  & Absolute magnitude (mag) \\
      5 & G                   & double  & Slope parameter \citep{1989-AsteroidsII-Bowell}\\
      6 & B\_V                & double  & B-V color (mag) from \citet{1989aste.conf.1090T}\\
      7 & IRAS\_diameter      & double  & IRAS diameter (km) from \citet{1989aste.conf.1151T} \\
      8 & IRAS\_class         & varchar & IRAS taxonomic classification from \citet{1989aste.conf.1151T}\\
      9 & note\_1             & int     & Categories of planet-crossing asteroids \\
      10& note\_2             & int     & Assumptions for orbit computation \\
      11& note\_3             & int     & Asteroids observed during the course of major surveys \\
      12& note\_4             & int     & Indication from MPC critical-list of numbered asteroids \\
      13& note\_5             & int     & Discoveries at Lowell Observatory and related discoveries \\
      14& note\_6             & int     & Rank for Lowell collaborative program of astrometry \\
      15& orbital\_arc        & int     & Orbital arc spanned by observations used in orbit computation (days) \\
      16& number\_observation & int     & Number of observations used in orbit computation \\
      17& yy\_osc             & int     & Year of the epoch of osculation \\
      18& mm\_osc             & int     & Month of the epoch of osculation \\
      19& dd\_osc             & int     & Day of the epoch of osculation \\
      20& mean\_anomaly       & double  & Mean anomaly (deg) \\
      21& perihelion\_argument& double  & Argument of perihelion (deg) in ECJ2000.0 \\
      22& node\_longitude     & double  & Longitude of ascending node (deg) in ECJ2000.0 \\
      23& inclination         & double  & Inclination (deg) in ECJ2000.0 \\
      24& eccentricity        & double  & Eccentricity  \\
      25& semi\_major\_axis   & double  & Semi-major axis (au) \\
      26& YY\_calculation     & int     & Year of the date of orbit computation \\
      27& MM\_calculation     & int     & Month of the date of orbit computation \\
      28& DD\_calculation     & int     & Day of the date of orbit computation \\
      29& CEU\_value          & double  & Absolute value of the Current 1-$\sigma$ Ephemeris Uncertainty (CEU, in arcsec) \\
      30& CEU\_rate           & double  & Rate of change of CEU (arcsec/day) \\
      31& CEU\_yy             & int     & Year of the date of CEU \\
      32& CEU\_mm             & int     & Month of the date of CEU \\
      33& CEU\_dd             & int     & Day of the date of CEU \\
      34& PEU\_value          & double  & Next Peak Ephemeris Uncertainty (PEU) from date of CEU (arcsec) \\
      35& PEU\_yy             & int     & Year of the date of occurrence of the PEU \\
      36& PEU\_mm             & int     & Month of the date of occurrence of the PEU \\
      37& PEU\_dd             & int     & Day of the date of occurrence of the PEU \\
      38& GPEU\_fromCEU       & double  & Greatest PEU in 10 years from date of CEU (arcsec) \\
      39& GPEU\_yy            & int     & Year of the date of occurrence of the GPEU \\
      40& GPEU\_mm            & int     & Month of the date of occurrence of the GPEU \\
      41& GPEU\_dd            & int     & Day of the date of occurrence of the GPEU \\
      42& GPEU\_fromPEU       & double  & Greatest PEU in 10 years from date of next PEU (arcsec) \\
      43& GGPEU\_yy           & int     & Year of the date of occurrence of the GPEU from PEU \\
      44& GGPEU\_mm           & int     & Month of the date of occurrence of the GPEU from PEU \\
      45& GGPEU\_dd           & int     & Day of the date of occurrence of the GPEU from PEU \\
      46& jd\_osc             & double  & JD of the epoch of osculation \\
      47& px                  & double  & x component of the EQJ2000 heliocentric position vector (au) \\
      48& py                  & double  & y component of the EQJ2000 heliocentric position vector (au) \\
      49& pz                  & double  & z component of the EQJ2000 heliocentric position vector (au) \\
      50& vx                  & double  & x component of the EQJ2000 heliocentric velocity vector (au/d) \\
      51& vy                  & double  & y component of the EQJ2000 heliocentric velocity vector (au/d) \\
      52& vz                  & double  & z component of the EQJ2000 heliocentric velocity vector (au/d) \\
      53& mean\_motion        & double  & Mean motion (deg/d) \\
      54& orbital\_period     & double  & Orbital period (d) \\
      55& iddataset           & int     & Bibliographic unique reference \\
   \end{longtable}
   
   \qquad

   \begin{longtable}{rlll}
      \caption{Description of the fields in the collection \colors of the \datacloud.\label{tab:struct:colors}}\\
      \hline\hline
      \# & Field & Type & Description \\
      \hline
      \endfirsthead
      \caption{continued.}\\
      \hline\hline
      \# & Field & Type & Description \\
      \hline
      \endhead
      \hline
      \endfoot
      1 & num           & int     & SSO IAU number \\   
      2 & name          & varchar & SSO name \\
      3 & color         & varchar & Name of the color (e.g, B-V)\\
      4 & value         & double  & Value of the color\\
      5 & uncertainty   & double  & Uncertainty on the color\\
      6 & facility      & varchar & Source of data (telescope, survey)\\
      7 & observer      & varchar & Observer IAU code\\
      8 & epoch         & double  & Epoch of observation (JD)\\
      9 & delta\_time   & float   & Time difference between filters (s)\\
      10& color\_type   & varchar & Description of the method (\Autoref{tab:methods}) \\
      11& id\_filter\_1 & varchar & First filter unique identifier  \citep[SVO filter service,][]{2012ivoa.rept.1015R}\\
      12& id\_filter\_2 & varchar & Second filter unique identifier \citep[SVO filter service,][]{2012ivoa.rept.1015R}\\
      13& phot\_sys     & varchar & Photometric system (Vega, AB, ST)\\
      14& selection     & int     & Selection flag (\Autoref{sec:ssocard}) \\
      15& iddataset     & int     & Bibliographic unique reference \\
   \end{longtable}

   \qquad

   \begin{longtable}{rlll}
      \caption{Description of the fields in the collection \cometpro of the \datacloud.\label{tab:struct:cometpro}}\\
      \hline\hline
      \# & Field & Type & Description \\
      \hline
      \endfirsthead
      \caption{continued.}\\
      \hline\hline
      \# & Field & Type & Description \\
      \hline
      \endhead
      \hline
      \endfoot
      \hline
      \multicolumn{4}{p{0.8\columnwidth}}{Fields follow the original COMETPRO data \citep{1996IAUS..172..357R}.}\\
      \endlastfoot
      2 & note                 & int     & Number of the note associated with the comet \\
      3 & updated              & date    & Date of update (DD/MM/YYYY) \\
      4 & name                 & varchar & IAU code of the comet \\
      5 & iau\_name            & varchar & IAU name of the comet \\
      6 & author               & varchar & Orbit computer \\
      7 & epoch                & double  & Reference epoch of the orbit (JD) \\
      8 & force\_relat         & int     & Relativity effect of the Sun taken into account (1) or not (0) \\
      9 & nb\_obs              & int     & Number of observations used in orbit computation \\
      10& sigma                & double  & 1-sigma residual (arcsec) \\
      11& start\_date          & date    & Date of first observation used in orbit computation (DD/MM/YYYY) \\
      12& end\_date            & date    & Date of last observation used in orbit computation (DD/MM/YYYY) \\
      13& px                   & double  & x component of the EQJ2000 heliocentric position vector (au) \\
      14& py                   & double  & y component of the EQJ2000 heliocentric position vector (au) \\
      15& pz                   & double  & z component of the EQJ2000 heliocentric position vector (au) \\
      16& vx                   & double  & x component of the EQJ2000 heliocentric velocity vector (au/d) \\
      17& vy                   & double  & y component of the EQJ2000 heliocentric velocity vector (au/d) \\
      18& vz                   & double  & z component of the EQJ2000 heliocentric velocity vector (au/d) \\
      19& fng\_A1              & double  & Radial non-gravitational acceleration (heliocentric EQJ2000) \\
      20& fng\_A2              & double  & Tangential non-gravitational acceleration (heliocentric EQJ2000)  \\
      21& fng\_A3              & double  & Normal non-gravitational acceleration (heliocentric EQJ2000) \\
      22& tau                  & double  & Date of perihelion passage (JD) \\
      23& perihelion\_distance & double  & Perihelion distance (au)\\
      24& eccentricity         & double  & Eccentricity\\
      25& perihelion\_argument & double  & Argument of perihelion (deg) (J2000.0)\\
      26& node\_longitude      & double  & Longitude of the ascending node (deg) (J2000.0)\\
      27& inclination          & double  & Inclination to ecliptic (deg) (J2000.0)\\
      28& mag\_H1              & double  & Constant term of magnitude to compute the total magnitude\\
      29& mag\_R1              & double  & Coefficient of log(r) to compute the total magnitude\\
      30& mag\_D1              & double  & Coefficient of log(Delta) to compute the total magnitude\\
      31& mag\_H2              & double  & Constant term of magnitude to compute the nuclear magnitude\\
      32& mag\_R2              & double  & Coefficient of log(r) to compute the nuclear magnitude\\
      33& mag\_D2              & double  & Coefficient of log(Delta) to compute the nuclear magnitude\\
      34& selection            & int     & Selection flag (\Autoref{sec:ssocard}) \\
      35& iddataset            & int     & Bibliographic unique reference \\
   \end{longtable}

   \clearpage

   \begin{longtable}{rlll}
      \caption{Description of the fields in the collection \density of the \datacloud.\label{tab:struct:density}}\\
      \hline\hline
      \# & Field & Type & Description \\
      \hline
      \endfirsthead
      \caption{continued.}\\
      \hline\hline
      \# & Field & Type & Description \\
      \hline
      \endhead
      \hline
      \endfoot
      1 & num                & int     & SSO IAU number \\      
      2 & name               & varchar & SSO name \\
      3 & density            & double  & Density in kg$\cdot $m$^{-3}$ \\
      4 & err\_density\_up   & double  & Upper uncertainty on the density (kg$\cdot $m$^{-3}$)\\
      5 & err\_density\_down & double  & Lower uncertainty on the density (kg$\cdot $m$^{-3}$)\\
      6 & method             & varchar & Description of the method (\Autoref{tab:methods}) \\
      7 & selection          & int     & Selection flag (\Autoref{sec:ssocard}) \\
      8 & iddataset          & int     & Bibliographic unique reference \\
   \end{longtable}

   \qquad

   \begin{longtable}{rlll}
      \caption{Description of the fields in the collection \diamalbedo of the \datacloud.\label{tab:struct:diam}}\\
      \hline\hline
      \# & Field & Type & Description \\
      \hline
      \endfirsthead
      \caption{continued.}\\
      \hline\hline
      \# & Field & Type & Description \\
      \hline
      \endhead
      \hline
      \endfoot
      1 & num                 & int     & SSO IAU number \\
      2 & name                & varchar & SSO name \\
      3 & diameter            & double  & Diameter in km \\
      4 & err\_diameter\_up   & double  & Upper uncertainty on the diameter (km) \\
      5 & err\_diameter\_down & double  & Lower uncertainty on the diameter (km) \\
      6 & albedo              & double  & Geometric visual albedo \\
      7 & err\_albedo\_up     & double  & Upper uncertainty on the albedo\\
      8 & err\_albedo\_down   & double  & Lower uncertainty on the albedo\\
      9 & beaming             & double  & Beaming parameter \citep{1999Icar..142..464H}\\
      10 & err\_beaming       & double  & Uncertainty on the beaming parameter\\
      11 & emissivity         & double  & Emissivity \citep{1999Icar..142..464H}\\
      12 & err\_emissivity    & double  & Uncertainty on the emissivity\\
      13 & selection          & int     & Selection flag (\Autoref{sec:ssocard}) \\
      14 & method             & varchar & Description of the method (\Autoref{tab:methods}) \\
      15 & iddataset          & int     & Bibliographic unique reference \\
   \end{longtable}

   \qquad

   \begin{longtable}{rlll}
      \caption{Description of the fields in the collection \family of the \datacloud.\label{tab:struct:fams}}\\
      \hline\hline
      \# & Field & Type & Description \\
      \hline
      \endfirsthead
      \caption{continued.}\\
      \hline\hline
      \# & Field & Type & Description \\
      \hline
      \endhead
      \hline
      \endfoot
      1 & num            & int     & SSO IAU number \\  
      2 & name           & varchar & SSO name \\
      3 & family\_status & varchar & SSO status: core, halo, diffuse halo\\
      4 & family\_num    & int     & IAU number of the family (if named after an asteroid)\\
      5 & family\_name   & varchar & Name of the family\\
      6 & selection      & int     & Selection flag (\Autoref{sec:ssocard}) \\
      7 & method         & varchar & Description of the method (\Autoref{tab:methods}) \\
      8 & iddataset      & int     & Bibliographic unique reference \\
   \end{longtable}
   
   \qquad

   \begin{longtable}{rlll}
      \caption{Description of the fields in the collection \masses of the \datacloud.\label{tab:struct:mass}}\\
      \hline\hline
      \# & Field & Type & Description \\
      \hline
      \endfirsthead
      \caption{continued.}\\
      \hline\hline
      \# & Field & Type & Description \\
      \hline
      \endhead
      \hline
      \endfoot
      1 & num             & int     & SSO IAU number \\ 
      2 & name            & varchar & SSO name \\
      3 & mass            & double  & Mass in kg \\
      4 & err\_mass\_up   & double  & Upper uncertainty on the mass (kg)\\
      5 & err\_mass\_down & double  & Lower uncertainty on the mass (kg)\\
      6 & method          & varchar & Description of the method (\Autoref{tab:methods}) \\
      7 & selection       & int     & Selection flag (\Autoref{sec:ssocard}) \\
      8 & method          & varchar & Description of the method (\Autoref{tab:methods}) \\
      9 & iddataset       & int     & Bibliographic unique reference \\
   \end{longtable}
   
   \qquad

   \begin{longtable}{rlll}
      \caption{Description of the fields in the collection \mpcatobs of the \datacloud.\label{tab:struct:mpcat}}\\
      \hline\hline
      \# & Field & Type & Description \\
      \hline
      \endfirsthead
      \caption{continued.}\\
      \hline\hline
      \# & Field & Type & Description \\
      \hline
      \endhead
      \hline
      \endfoot
      \hline
      \multicolumn{4}{p{0.8\columnwidth}}{MPC Web site: \url{https://minorplanetcenter.net/iau/info/OpticalObs.html}}\\
      \endlastfoot
      1 & type               & varchar  & Type of SSO (asteroid, comet)\\
      2 & num                & varchar  & SSO number\\
      3 & packed\_name       & varchar  & SSO packed name\\
      4 & name               & varchar  & SSO name\\
      5 & orbit\_type        & varchar  & Type of orbit (for comets)\\
      6 & discovery          & varchar  & Discovery asterisk\\
      7 & note1              & varchar  & See MPC Web site\\
      8 & note2              & varchar  & See MPC Web site\\
      9 & date\_obs          & datetime & Date of observation (ISO)\\
      10& jd\_obs            & double   & Date of observation (JD)\\
      11& ra\_obs            & double   & Observed right ascension (deg) (EQJ2000.0)\\
      12& dec\_obs           & double   & Observed declination (deg) (EQJ2000.0)\\
      13& mag                & double   & Observed magnitude (mag)\\
      14& filter             & varchar  & Magnitude band\\
      15& astrocata\_name    & varchar  & Astrometric reference catalog used to determine the position \\
      16& astrocata\_vizname & varchar  & VizieR table base-name of the astrometric reference catalog \\
      17& mpc\_ref           & varchar  & Permanent references to the MPCs, MPSs, or other journals \\
      18& iau\_code          & varchar  & IAU observatory code\\
      19& obs\_long          & double   & Geographic longitude of observing site (deg) \\
      20& obs\_lat           & double   & Geographic latitude of observing site (deg) \\
      21& obs\_alt           & double   & Altitude of observing site (m) \\
      22& vgs\_x             & double   & x component of spacecraft geocentric position vector (au) (EQJ2000.0)\\
      23& vgs\_y             & double   & y component of spacecraft geocentric position vector (au) (EQJ2000.0)\\
      24& vgs\_z             & double   & z component of spacecraft geocentric position vector (au) (EQJ2000.0)\\
      25& iddataset          & int      & Bibliographic unique reference \\
   \end{longtable}

   \qquad

   \begin{longtable}{rlll}
      \caption{Description of the fields in the collection \mpcorb of the \datacloud.\label{tab:struct:mpcorb}}\\
      \hline\hline
      \# & Field & Type & Description \\
      \hline
      \endfirsthead
      \caption{continued.}\\
      \hline\hline
      \# & Field & Type & Description \\
      \hline
      \endhead
      \hline
      \endfoot
      \hline
      \multicolumn{4}{p{0.9\columnwidth}}{Fields follows the original MPCORB data, see the online documentation \url{https://www.minorplanetcenter.net/iau/info/MPOrbitFormat.html}.}\\
      \endlastfoot
      1 & packed\_name        & varchar  & Packed number or name of the SSO \\ 
      2 & num                 & int      & SSO IAU Number \\
      3 & name                & varchar  & SSO name \\ 
      4 & H                   & double   & Absolute magnitude (mag) \\
      5 & G                   & double   & Slope parameter \citep{1989-AsteroidsII-Bowell}\\
      6 & ref\_date           & datetime & Reference epoch TT (ISO) \\
      7 & mean\_anomaly       & double   & Mean anomaly (deg) \\
      8 & perihelion\_argument& double   & Argument of perihelion (deg) in ECJ2000.0 \\
      9 & node\_longitude     & double   & Longitude of ascending node (deg) ECJ2000.0 \\
      10& inclination         & double   & Inclination (deg) in ECJ2000.0 \\
      11& eccentricity        & double   & Eccentricity  \\
      12& mean\_motion        & double   & Mean motion (deg/d) \\
      13& semi\_major\_axis   & double   & Semi-major axis (au) \\
      14& U                   & varchar  & Uncertainty parameter \\
      15& reference           & varchar  & Orbit reference \\
      16& number\_observation & int      & Number of observations used to compute the orbit \\
      17& number\_opposition  & int      & Number of oppositions \\
      18& start\_obs          & int      & Year of the first observation \\
      19& end\_obs            & int      & Year of the last observation \\
      20& orbital\_arc        & double   & Orbit arc length (d) \\
      21& rms                 & double   & Root-mean square residuals of the fit (arcsec) \\
      22& coarse\_indic       & varchar  & Coarse indicator of perturbers \\
      23& precise\_indic      & varchar  & Precise indicator of perturbers \\
      24& orbit\_computer     & varchar  & Orbit computer \\
      25& orbit\_type         & varchar  & 4-hexdigit flags describing the orbit \\
      26& last\_obs           & double   & Date of last observation included in orbit solution (YYYYMMDD) \\
      27& jd\_osc             & double   & JD of the epoch of osculation \\
      28& px                  & double   & x component of heliocentric position vector (au, EQJ2000)\\
      29& py                  & double   & y component of heliocentric position vector (au, EQJ2000)\\
      30& pz                  & double   & z component of heliocentric position vector (au, EQJ2000)\\
      31& vx                  & double   & x component of heliocentric velocity vector (au/d, EQJ2000)\\
      32& vy                  & double   & y component of heliocentric velocity vector (au/d, EQJ2000)\\
      33& vz                  & double   & z component of heliocentric velocity vector (au/d, EQJ2000)\\
      34& orbital\_period     & double   & Orbital period (d) \\
      35& iddataset           & int      & Bibliographic unique reference \\
   \end{longtable}

   \qquad

   \begin{longtable}{rlll}
      \caption{Description of the fields in the collection \pairs of the \datacloud.\label{tab:struct:pairs}}\\
      \hline\hline
      \# & Field & Type & Description \\
      \hline
      \endfirsthead
      \caption{continued.}\\
      \hline\hline
      \# & Field & Type & Description \\
      \hline
      \endhead
      \hline
      \endfoot
      1 & num           & int     & First member IAU number \\  
      2 & name          & varchar & First member name \\
      3 & sibling\_num  & int     & Second member IAU number \\ 
      2 & sibling\_name & varchar & Second member name \\
      3 & distance      & double  & Orbital distance (m/s)\\
      4 & age           & double  & Estimated age of the pair (kyr)\\
      4 & err\_age\_up  & double  & Upper uncertainty on the age (kyr)\\
      5 & err\_age\_down& double  & Lower uncertainty on the age (kyr)\\
      7 & selection     & int     & Selection flag (\Autoref{sec:ssocard}) \\
      6 & method        & varchar & Description of the method (\Autoref{tab:methods}) \\
      8 & iddataset     & int     & Bibliographic unique reference \\
   \end{longtable}

   \qquad

   \begin{longtable}{rlll}
      \caption{Description of the fields in the collection \phase of the \datacloud.\label{tab:struct:phase}}\\
      \hline\hline
      \# & Field & Type & Description \\
      \hline
      \endfirsthead
      \caption{continued.}\\
      \hline\hline
      \# & Field & Type & Description \\
      \hline
      \endhead
      \hline
      \endfoot
      1 & num           & int     & SSO IAU Number \\   
      2 & name          & varchar & SSO name \\
      3 & H             & double  & Absolute magnitude\\
      4 & G1            & double  & Phase parameter G$_1$ \citep{2010Icar..209..542M}\\
      5 & G2            & double  & Phase parameter G$_2$ \citep{2010Icar..209..542M}\\
      6 & err\_H\_down  & double  & Lower uncertainty on absolute magnitude\\
      7 & err\_H\_up    & double  & Upper uncertainty on absolute magnitude\\
      8 & err\_G1\_down & double  & Lower uncertainty on G$_1$ phase parameter\\
      9 & err\_G1\_up   & double  & Upper uncertainty on G$_1$ phase parameter\\
      10& err\_G2\_down & double  & Lower uncertainty on G$_2$ phase parameter\\
      11& err\_G2\_up   & double  & Upper uncertainty on G$_2$ phase parameter\\
      12& N             & double  & Number of observations used to derive (H,G$_1$,G$_2$)\\
      13& phase\_min    & double  & Minimum phase angle (\degr)\\
      14& phase\_max    & double  & Maximum phase angle (\degr)\\
      15& rms           & double  & Root mean-square of the fit (mag)\\
      16& facility      & varchar & Source of observations (telescope, survey)\\
      17& name\_filter  & varchar & Name of the filter\\
      18& id\_filter    & varchar & Filter unique identifier  \citep[SVO filter service,][]{2012ivoa.rept.1015R}\\
      19& method        & varchar & Description of the method (\Autoref{tab:methods}) \\
      20& selection     & int     & Selection flag (\Autoref{sec:ssocard}) \\
      21& iddataset     & int     & Bibliographic unique reference \\
   \end{longtable}

   \qquad
   
   \begin{longtable}{rllR}
      \caption{Description of the fields in the collection \proper of the \datacloud.\label{tab:struct:proper}}\\
      \hline\hline
      \# & Field & Type & Description \\
      \hline
      \endfirsthead
      \caption{continued.}\\
      \hline\hline
      \# & Field & Type & Description \\
      \hline
      \endhead
      \hline
      \endfoot
      1 & num                                             & int     & SSO IAU Number \\   
      2 & name                                            & varchar & SSO name \\
      3 & H                                               & double  & Absolute magnitude \\
      4 & proper\_semi\_major\_axis                       & double  & Proper semi-major axis (au) \\
      5 & err\_proper\_semi\_major\_axis                  & double  & Uncertainty on proper semi-major axis (au) \\
      6 & proper\_eccentricity                            & double  & Proper eccentricity \\
      7 & err\_proper\_eccentricity                       & double  & Uncertainty on proper eccentricity \\
      8 & proper\_sine\_inclination                       & double  & Sine of proper inclination \\
      9 & err\_proper\_sine\_inclination                  & double  & Uncertainty on sine of proper inclination \\
      10 & proper\_inclination                            & double  & Proper inclination (\degr) \\
      11 & err\_proper\_inclination                       & double  & Uncertainty on proper inclination (\degr) \\
      12 & proper\_frequency\_mean\_motion                & double  & Proper frequency of mean motion (\degr/yr) \\
      13 & err\_proper\_frequency\_mean\_motion           & double  & Uncertainty on proper frequency of mean motion (\degr/yr) \\
      14 & proper\_frequency\_perihelion\_longitude       & double  & Proper frequency of perihelion longitude (arcsec/yr) \\
      15 & err\_proper\_frequency\_perihelion\_longitude  & double  & Uncertainty on proper frequency of perihelion longitude (arcsec/yr) \\
      16 & proper\_frequency\_nodal\_longitude            & double  & Proper frequency of nodal longitude (arcsec/yr) \\
      17 & err\_proper\_frequency\_nodal\_longitude       & double  & Uncertainty on proper frequency of nodal longitude (arcsec/yr) \\
      18 & lyapunov\_time                                 & double  & Timescale of chaoticity (yr) \\
      19 & integration\_time                              & double  & Length of integration (Myr) \\
      20 & identfrom                                      & varchar & Name of the SSO in the imported data \\
      21 & iddataset                                      & int     & Bibliographic unique reference \\
   \end{longtable}

   \qquad

   \begin{longtable}{rlll}
      \caption{Description of the fields in the collection \spin of the \datacloud.\label{tab:struct:spin}}\\
      \hline\hline
      \# & Field & Type & Description \\
      \hline
      \endfirsthead
      \caption{continued.}\\
      \hline\hline
      \# & Field & Type & Description \\
      \hline
      \endhead
      \hline
      \endfoot
      1  & num             & int     & SSO IAU number \\        
      2  & name            & varchar & SSO name \\
      3  & model\_name     & varchar & Name of the model\\
      4  & t0              & double  & Reference epoch for spin coordinates (JD)\\
      5  & W0              & double  & Rotation phase at t0 \citep[\degr,][]{2018CeMDA.130...22A} \\
      6  & Wp              & double  & Rotation velocity \citep[\degr/d,][]{2018CeMDA.130...22A} \\
      7  & RA0             & double  & EQJ2000 right ascencion of the spin axis (\degr)  \\
      8  & DEC0            & double  & EQJ2000 declination of the spin axis (\degr) \\
      9  & err\_RA0        & double  & Uncertainty on the right ascencion (\degr)  \\
      10 & err\_DEC0       & double  & Uncertainty on the declination (\degr) \\
      11 & period          & double  & Rotation period (h) \\
      12 & err\_period     & double  & Uncertainty on rotation period (h) \\
      13 & period\_flag    & double  & Rotation period quality code \citep{2021pdss.data...10W} \\
      14 & period\_type    & varchar & Sidereal or synodic \\
      15 & long            & double  & ECJ2000 longitude of the spin axis (\degr)  \\
      16 & lat             & double  & ECJ2000 latitude of the spin axis (\degr) \\
      17 & err\_long       & double  & Uncertainty on the longitude (\degr)  \\
      18 & err\_lat        & double  & Uncertainty on the latitude (\degr) \\
      19 & selection       & int     & Selection flag (\Autoref{sec:ssocard}) \\
      20 & method          & varchar & Description of the method (\Autoref{tab:methods}) \\
      21 & iddataset       & int     & Bibliographic unique reference \\
   \end{longtable}
 
   \qquad
 
   \begin{longtable}{rlll}
      \caption{Description of the fields in the collection \taxonomy of the \datacloud.\label{tab:struct:taxo}}\\
      \hline\hline
      \# & Field & Type & Description \\
      \hline
      \endfirsthead
      \caption{continued.}\\
      \hline\hline
      \# & Field & Type & Description \\
      \hline
      \endhead
      \hline
      \endfoot
      1 & num             & int     & SSO IAU number \\ 
      2 & name            & varchar & SSO name \\
      3 & scheme          & varchar & Taxonomic scheme (e.g., Tholen, Bus, DeMeo, Mahlke) \\
      4 & class           & varchar & Taxonomic class \\
      5 & complex         & varchar & Taxonomic complex (\Autoref{tab:taxo}) \\
      6 & selection       & int     & Selection flag (\Autoref{sec:ssocard}) \\
      7 & method          & varchar & Description of the method (\Autoref{tab:methods}) \\
      8 & waverange       & varchar & Waverange used in taxonomy (VIS, NIR, VISNIR) \\
      9 & iddataset       & int     & Bibliographic unique reference \\
   \end{longtable}
   
   \qquad

   \begin{longtable}{rlll}
      \caption{Description of the fields in the collection \thermal of the \datacloud.\label{tab:struct:thermal}}\\
      \hline\hline
      \# & Field & Type & Description \\
      \hline
      \endfirsthead
      \caption{continued.}\\
      \hline\hline
      \# & Field & Type & Description \\
      \hline
      \endhead
      \hline
      \endfoot
      1 & num             & int     & SSO IAU number \\ 
      2 & name            & varchar & SSO name \\
      3 & TI              & double  & Thermal inertia (J.s$^{-1/2}$K$^{-1}$m$^{-2}$) \\
      4 & err\_TI\_up     & double  & Upper uncertainty on the thermal inertia\\
      5 & err\_TI\_down   & double  & Lower uncertainty on the thermal inertia\\
      6 & dsun            & double  & Heliocentric distance at the time of measurements (au)\\
      7 & selection       & int     & Selection flag (\Autoref{sec:ssocard}) \\
      8 & method          & varchar & Description of the method (\Autoref{tab:methods}) \\
      9 & iddataset       & int     & Bibliographic unique reference \\
   \end{longtable}
   
   \qquad

   \begin{longtable}{rlll}
      \caption{Description of the fields in the collection \yarko of the \datacloud.\label{tab:struct:yarko}}\\
      \hline\hline
      \# & Field & Type & Description \\
      \hline
      \endfirsthead
      \caption{continued.}\\
      \hline\hline
      \# & Field & Type & Description \\
      \hline
      \endhead
      \hline
      \endfoot
      1  & num       & int     & SSO IAU Number \\      
      2  & name      & varchar & SSO name \\
      3  & A2        & double  & Radial acceleration ($10^{-15}$ au/$d^2$)\\
      4  & err\_A2   & double  & Uncertainty on radial acceleration ($10^{-15}$ au/$d^2$)\\
      5  & dadt      & double  & Semi-major drift ($10^{-4}$ au/Myr) \\
      6  & err\_dadt & varchar & Uncertainty on semi-major drift ($10^{-4}$ au/Myr) \\ 
      7  & snr       & float   & Signal-to-noise ratio \\ 
      8  & S         & float   & Sensitivity parameter \citep{2012AJ....144...60N} \\
      9  & selection & int     & Selection flag (\Autoref{sec:ssocard}) \\
      10 & method    & varchar & Description of the method (\Autoref{tab:methods}) \\
      11 & iddataset & int     & Bibliographic unique reference \\
   \end{longtable}

   \clearpage
   \section{Description of all the methods\label{app:method}}
   
   \begin{longtable}{p{2.5cm}MNL}
      \caption{Methods included in \ssodnet.\label{tab:methods}}\\
      \hline\hline
      Method & Name & Description & Reference \\
      \hline
      \endfirsthead
      \caption{continued.}\\
      \hline\hline
      Method & Name & Description & Reference \\
      \hline
      \endhead
      \hline
      \endfoot
      {\texttt{SPACE}}& Rendez-vous  with  a  spacecraft & The  results  are  based  on  data  which  had  an  encounter  (flyby  or  orbit)  with  the  target & {\citet{1992Sci...257.1647B}}  \\ 
      {\texttt{STM}}& Standard  Thermal  Model & Diameter  and  albedo  derived  by  fitting  mid-infrared  data  with  a  simple  thermal  model  of  non-rotating  spheres & {\citet{1986Icar...68..239L}}  \\ 
      {\texttt{NEATM}}& Near-Earth  Asteroid  Thermal  Model & Diameter,  albedo,  beaming  derived  by  fitting  mid-infrared  data  with  a  simple  thermal  model & {\citet{1999Icar..142..464H}}  \\ 
      {\texttt{TPM}}& ThermoPhysical  Model & Diameter,  albedo,  thermal  inertia  derived  by  fitting  mid-infrared  data  with  a  thermal  model  taking  into  account  the  spin,  shape  of  the  target & {\citet{1996A&A...310.1011L}}  \\ 
      {\texttt{PhaseFunction}}& Albedo  determined  from  the  phase  function & Albedo  determined  from  the  phase  function & {\citet{2000Icar..147...94B}}  \\ 
      {\texttt{LC}}&Lightcurve& Rotation  period  determined  from  optical  light curves & {\citet{1932AN....246..441Z}}  \\ 
      {\texttt{Comet-Break}}& Mass  from  break-up & Mass  estimated  from  the  break-up  of  the  comet & {\citet{1994Natur.370..349S}}  \\ 
      {\texttt{FRM}}& Fast  Rotating  Model & Diameter  and  albedo  derived  by  fitting  mid-infrared  data  with  a  simple  thermal  model  of  rapidly  non-rotating  spheres & {\citet{1989aste.conf..128L}}  \\ 
      {\texttt{NESTM}}& Night  Emission  Simulated  Thermal  Model & Diameter,  albedo,  beaming  derived  by  fitting  mid-infrared  data  with  an  adapted  NEATM & {\citet{2009MNRAS.400..204W}}  \\ 
      {\texttt{Speckle}}& Triaxial  ellipsoid  from  speckle  interferometry & 3D  shape  modeled  as  tri-axial  ellipsoid  using  speckle  interferometry & {\citet{1985Icar...61..132D}}  \\ 
      {\texttt{Interferometry}}& Optical  and  Infrared  Interferometry & Diameter  derived  from  interferometric  visibilities  in  the  optical  or  infrared & {\citet{2009ApJ...694.1228D}}  \\ 
      {\texttt{Occ}}& Stellar  Occultation & Apparent  size  measured  during  a  stellar  occultation & {\citet{1979RMxAA...4..205D}}  \\ 
      {\texttt{IM}}& Apparent  shape  from  direct  imaging & Apparent  size/shape  measured  on  disk-resolved  images & {\citet{2006Icar..185...39M}}  \\ 
      {\texttt{IM-PSF}}& Diameter  from  PSF  deviation & Estimate  of  diameter  from  the  deviation  of  the  PSF  compared  with  a  star & {\citet{2004AJ....127.2413B}}  \\ 
      {\texttt{TE-IM}}& Triaxial  ellipsoid  from  disk-resolved  imaging & 3D  shape  modeled  as  tri-axial  ellipsoid  using  disk-resolved  images & {\citet{2000ASIC..551..243D}}  \\ 
      {\texttt{TE-Occ}}& Triaxial  ellipsoid  from  stellar  occultation & 3D  shape  modeled  as  tri-axial  ellipsoid  using  stellar  occultations & {\citet{1989Icar...78..323D}}  \\ 
      {\texttt{ADAM}}& All-Data  Asteroid  Model & 3D  shape  model  obtained  from  a  combined  use  of  stellar  occultations,  optical  light curves,  disk-resolved  images,  interferometric  fringes & {\citet{2015A&A...576A...8V}}  \\ 
      {\texttt{KOALA}}& Knitted  Occultation,  Adaptive-optics,  and  Lightcurves  Analysis & The  results  are  obtained  from  the  combined  use  of  stellar  occultation,  optical  light curves,  and  disk-resolved  images & {\citet{2010Icar..205..460C}}  \\ 
      {\texttt{Radar}}& Radar  shape  modeling & 3D  shape  model  based  on  radar  Delay-Doppler  data & {\citet{1994Sci...263..940H}}  \\ 
      {\texttt{Radar-LC}}& Combined  radar  and  light curve  shape  modeling & 3D  shape  model  based  on  radar  Delay-Doppler  and  optical  light curve  data & {\citet{1997Icar..130..165H}}  \\ 
      {\texttt{SAGE}}& Shaping  Asteroids  with  Genetic  Evolution & 3D  shape  model  based  on  light curves,  found  by  genetic  evolution & {\citet{2018MNRAS.473.5050B}}  \\ 
      {\texttt{Polarimetry}}& Albedo  determined  from  polarimetry & Albedo  determined  from  polarimetry & {\citet{1999Icar..138..129C}}  \\ 
      {\texttt{A-M}}&Amplitude-Magnitude& Determination  of  the  spin  axis  from  the  amplitude  of  light curves & {\citet{1983acm..proc...73Z}}  \\ 
      {\texttt{TE}}& Triaxial ellipsoid from light curves & Determination of the spin axis, modeling the light curves with a triaxial ellipsoid & {\citet{2021A&A...654A..48H}}  \\ 
      {\texttt{LCI}}& Lightcurve  Inversion & Spin  and  convex  3-D  shape  determined  from  optical  light curves & {\citet{2001Icar..153...24K}}  \\ 
      {\texttt{LC+Occ}}& Scaling  of  Lightcurve  Inversion  Model  with  Stellar  Occultations & 3D  shape  model  from  light-curve  inversion  scaled  using  stellar  occultation(s) & {\citet{2011Icar..214..652D}}  \\ 
      {\texttt{LC+IM}}& Scaling  of  Lightcurve  Inversion  Model  with  direct imaging & 3D  shape  model  from  a light-curve  inversion  scaled  using  disk-resolved  image(s) & {\citet{2013Icar..226.1045H}}  \\ 
      {\texttt{LC+TPM}}& Scaling  of  light-curve  inversion  model  with the thermophysical  model & 3D  shape  model  from  a light-curve  inversion  scaled  using  a  thermophysical  model  on  mid-infrared  data & {\citet{2015Icar..256..101H}}  \\ 
      {\texttt{LC-TPM}}& Combined  light-curve  inversion  and  thermophysical  modeling & 3D  shape  modeling  from  simultaneous  a light-curve  inversion  and  thermophysical  model  of  mid-infrared  data & {\citet{2017A&A...604A..27D}}  \\ 
      {\texttt{EPHEM}}& Mass  from  ephemerides & The  mass  is  determined  from  general  ephemerides  of  the  Solar  System & {\citet{2008CeMDA.100...27B}}  \\ 
      {\texttt{DEFLECT}}& Mass  from  close  encounter  deflection & The  mass  is  determined  from  the  orbital  deflection  of  smaller  asteroids & {\citet{1989Icar...80..326S}}  \\ 
      {\texttt{Bin-IM}}& Mass  from  optical  imaging  a  binary  system & Mass  from  a  binary  system  imaged  in  the  optical & {\citet{1999Natur.401..565M}}  \\ 
      {\texttt{Bin-Radar}}& Mass  from  radar  observations  of  a  binary  system & Mass  from  a  binary  system  observed  by  radar  echoes & {\citet{2006Sci...314.1276O}}  \\ 
      {\texttt{Bin-PheMu}}& Mass  from  mutual  phenomena  in  a  binary  system & Mass  from  a  binary  system  from  the  timings and shape  of  mutual  event  from  light curves & {\citet{2000Icar..146..190P}}  \\ 
      {\texttt{Bin-Genoid}}& Orbit  and  mass  from  a  multiple  asteroidal  system  using  Genoid  algorithm & Orbital  elements  and  mass  determination  from  a  multiple  asteroidal  system  with  Genoid & {\citet{2012A&A...543A..68V}}  \\ 
      {\texttt{Yarkovsky}}& Mass  from  Yarkovsky  drift & Determination  of  the  mass  from  the  measured  Yarkovsky  drift & {\citet{2014Icar..235....5C}}  \\ 
      {\texttt{Comet-NGF}}& Mass  from  non-gravitational  forces & Mass  estimated  from  the  non-gravitational  acceleration & {\citet{2007Icar..187..306D}}  \\ 
      {\texttt{Spec}}& Reflectance  spectroscopy & Reflectance  spectroscopy & {\citet{1970Sci...168.1445M}}  \\ 
      {\texttt{Phot}}& Multi-filter  photometry & Multi-band  photometry & {\citet{2013Icar..226..723D}}  \\ 
      {\texttt{Astrometry(O)}}& Yarkovsky  drift  from  optical  astrometry & Determination  of  the  semi-major  drift  due  to  Yarkovsky  using  astrometry  from  optical  observations & {\citet{2003Sci...302.1739C}}  \\ 
      {\texttt{Astrometry(O+R)}}& Yarkovsky  drift  from  optical  astrometry  and  radar  delays & Determination  of  the  semi-major  drift  due  to  Yarkovsky  using  astrometry  from  optical  observations  and  radar  delays & {\citet{2003Sci...302.1739C}}  \\ 
      {\texttt{Family\_age}}& Yarkovsky  drift  from  family  age & Determination  of  the  semi-major  drift  due  to  Yarkovsky  using  the  age  of  the  dynamical  family & {\citet{2017MNRAS.469.4400C}}  \\ 
      {\texttt{HCM}}& Hierarchical  Clustering  Method & Determination  of  family  membership  by  hierarchical  clustering  of  proper  elements & {\citet{1990AJ....100.2030Z}}  \\ 
      {\texttt{V-Shape}}& Yarkovsky  V-shape  identification  of  asteroid  families & Determination  of  family  membership  by  identification  of  the Yarkovsky  print  in  (semi-major  axis,  1/diameter)  plane & {\citet{2017Icar..282..290B}}  \\ 
      {\texttt{abs}}& Colors  derived  from  absolute  magnitudes & Colors  computed  from  the  absolute  magnitudes  in  the  two  filters & {\citet{2021Icar..35414094M}}  \\ 
      {\texttt{lc\_cor}}& Colors  derived  from  apparent  magnitudes  corrected  for  light curves & Colors  computed  from  the  apparent  magnitudes,  corrected  for  short-term  variability  introduced  by  light curves & {\citet{2019ApJS..242...15E}}  \\ 
      {\texttt{app}}& Colors  derived  from  apparent  magnitudes & Colors  computed  from  the  apparent  magnitudes & {\citet{2000Icar..146..161S}}  \\ 
      {\texttt{Yarkovsky\_drift}}& Thermal  inertia  derived  from  Yarkovsky  drift & Determination  of  the  thermal  inertia  based  on  the  measured  strength  of  the  Yarkovsky  effect & {\citet{2021A&A...647A..61F}}  \\ 
      {\texttt{serendipitous}}& Phase curve from serendipitous observations & Determination of the parameters of the phase function from serendipitous observations (from surveys) & {\citet{2011JQSRT.112.1919O}}  \\ 
      {\texttt{targeted}}& Phase curve from targeted observations & Determination of the parameters of the phase function from targeted observations (generally a reduction to the light curve maxima) & {\citet{1956ApJ...123..331G}}  \\ 
   \end{longtable}

   \clearpage
   \section{Method lists for best-estimate determination\label{app:order}}
   
   \begin{table}[h]
     \caption{Ranking of methods for diameter estimates (\diamalbedo).\label{tab:rank:diam}}
     \begin{tabular}{p{0.1\linewidth}p{0.8\linewidth}}
       \hline\hline
       Order & Methods  \\
       \hline
       1 & \texttt{SPACE} \\
       2 & \texttt{ADAM, KOALA, SAGE, Radar} \\
       3 & \texttt{LC+Occ, LC+IM, LC+TPM, TPM, TE-IM, TE-Occ} \\ 
       4 & \texttt{IM, Occ, IM-PSF, Interferometry} \\ 
       5 & \texttt{NEATM, NESTM} \\ 
       6 & \texttt{STM, FRM} \\ 
       7 & \texttt{Polarimetry} \\ 
       \hline
       \multicolumn{2}{p{0.9\columnwidth}}{The order favors direct measurements first, then estimates based on 3D shape models,
          followed by direct measurements limited to a single geometry, and, finally, indirect estimates from thermal
          model of spheres.}
     \end{tabular}
   \end{table}

   \begin{table}[h]
     \caption{Selection order for albedo determinations (\diamalbedo).\label{tab:rank:albedo}}
     \begin{tabular}{p{0.1\linewidth}p{0.8\linewidth}}
       \hline\hline
       Order & Methods  \\
       \hline
       1 & \texttt{SPACE} \\
       2 & \texttt{Polarimetry} \\ 
       \hline
     \end{tabular}
   \end{table}

   \begin{table}[h]
     \caption{Selection order for mass determinations (\masses).\label{tab:rank:mass}}
     \begin{tabular}{p{0.1\linewidth}p{0.8\linewidth}}
       \hline\hline
       Order & Methods  \\
       \hline
       1 & \texttt{SPACE} \\
       2 & \texttt{Bin-Genoid} \\
       3 & \texttt{Bin-IM, Bin-Radar, Bin-PheMu} \\ 
       4 & \texttt{Deflect, Ephem} \\
       5 & \texttt{Yarkovsky} \\
       \hline
       \multicolumn{2}{p{0.9\columnwidth}}{The order favors spacecraft encounters, followed by binary systems, and, finally, estimates based on long-distance gravitational interactions and Yarkovsky drift.}
     \end{tabular}
   \end{table}

   \begin{table}[h]
     \caption{Ranking of methods for spin properties (\spin).\label{tab:rank:spin}}
     \begin{tabular}{p{0.1\linewidth}p{0.8\linewidth}}
       \hline\hline
       Order & Methods  \\
       \hline
       1 & \texttt{SPACE} \\
       2 & \texttt{ADAM, KOALA, SAGE, Radar, Radar-LC} \\
       3 & \texttt{LC+TPM, LC-TPM, LC+IM, LC+Occ} \\
       4 & \texttt{LCI} \\
       5 & \texttt{LC, A-M, Bin-IM, TE, TE-IM, TE-Occ, Speckle} \\
       \hline
       \multicolumn{2}{p{0.9\columnwidth}}{The order favors solutions from spacecrafts encounters, 
          followed by 3D shape modeling, ranked from modeling including direct measurement to scaling 
          of 3D convex models, to convex model of arbitrary size. Simple ellipsoids follow and, finally, the 
          periods from light curves.}
     \end{tabular}
   \end{table}

   \begin{table}[h!]
     \caption{Selection order for colors (\colors).\label{tab:rank:color}}
     \begin{tabular}{p{0.1\linewidth}p{0.8\linewidth}}
       \hline\hline
       Order & Methods  \\
       \hline
       1 & \texttt{Absolute} \\
       2 & \texttt{Light-curve corrected} \\
       3 & \texttt{Apparent} \\
       \hline
       \multicolumn{2}{p{0.9\columnwidth}}{The order favors absolute magnitude over
          light-curve corrected magnitude, and, finally, over apparent magnitude.}
     \end{tabular}
   \end{table}

   \clearpage

   \begin{table}[h]
     \caption{Ranking of methods for thermal properties (\thermal).\label{tab:rank:thermal}}
     \begin{tabular}{p{0.1\linewidth}p{0.8\linewidth}}
       \hline\hline
       Order & Methods  \\
       \hline
       1 & \texttt{SPACE} \\
       2 & \texttt{LCI-TPM, LCI+TPM} \\
       3 & \texttt{TPM} \\
       4 & \texttt{Yarkovsky\_drift} \\
       \hline
       \multicolumn{2}{p{0.9\columnwidth}}{The order favors estimates from space mission encounters,
          then thermophysical models based on 3D shapes, and, finally, estimates from thermophysical models 
          based on limited shape and spin information.}
     \end{tabular}
   \end{table}

   \begin{table}[h]
     \caption{Selection order for Yarkovsky drift determinations (\yarko).\label{tab:rank:yarko}}
       \begin{tabular}{p{0.1\linewidth}p{0.8\linewidth}}
       \hline\hline
       Order & Methods  \\
       \hline
       1 & \texttt{Astrometry(O+R)} \\
       2 & \texttt{Astrometry(O)} \\
       3 & \texttt{Family\_age} \\
       \hline
       \multicolumn{2}{p{0.9\columnwidth}}{The order favors solutions using a combination of optical 
          and radar observations over optical-only data sets, and, finally, estimates based on family ages.}
     \end{tabular}
   \end{table}

   \begin{table}[h!]
      \caption{Class-complex connections for taxonomy.\label{tab:taxo}}
      \begin{tabular}{clp{0.6\columnwidth}}
         \hline\hline
         Complex & Reference & Classes \\
         \hline
         \texttt{A}  & \citet{1983Icar...55..177V} & \texttt{A} \\
         \texttt{B}  & \citet{1984PhDT.........3T} & \texttt{B} \\
         \texttt{C}  & \citet{1975Icar...25..104C} & \texttt{C, Cb, CF, CFB, Cg, CG, Cgx, F, FC, G, GC} \\
         \texttt{Ch} & \citet{2002Icar..158..146B} & \texttt{Caa, Cgh, Ch} \\
         \texttt{D}  & \citet{1982Sci...216.1405G} & \texttt{D} \\
         \texttt{E}  & \citet{1976AJ.....81..262Z} & \texttt{E} \\
         \texttt{K}  & \citet{1989aste.conf.1151T} & \texttt{K} \\
         \texttt{L}  & \citet{1988LPI....19...57B} & \texttt{L, Ld} \\
         \texttt{M}  & \citet{1976AJ.....81..262Z} & \texttt{M} \\
         \texttt{O}  & \citet{1993Sci...262.1541B} & \texttt{O} \\
         \texttt{P}  & \citet{1982Sci...216.1405G} & \texttt{DP, P, PD} \\
         \texttt{Q}  & \citet{1984PhDT.........3T} & \texttt{Q, QO, Qw} \\
         \texttt{R}  & \citet{1978Icar...35..313B} & \texttt{R} \\
         \texttt{S}  & \citet{1975Icar...25..104C} & \texttt{S, SA, Sa, Sk, Sl, SO, Sq, SQ, Sqw, Sr, SR, Srw, SV, Sv, Svw, Sw} \\
         \texttt{T}  & \citet{1977cami.coll..185Z} & \texttt{T} \\
         \texttt{V}  & \citet{1970Sci...168.1445M} & \texttt{J, V, Vw} \\
         \texttt{X}  & \citet{1984PhDT.........3T} & \texttt{EM, X, Xc, Xe, Xk, Xn, Xt} \\
         \texttt{Z}  & \citet{1992Icar...97..150M} & \texttt{Z} \\
         \texttt{U}  & & \texttt{Ad, AQ, AS, AU, AV, BC, BCF, BCU, BFC, BFU, BFX, Bk, BU, CB, CBU, CD, CDX, CFU, CFXU, CGSU, CGTP, CGU, CL, CO, CP, CPF, CPU, CQ, CS, CSGU, CSU, CTGU, CU, CX, CXF, DCX, DL, Ds, DS, DSU, DT, DTU, DU, DX, DXCU, EU, FBCU, FCB, FCU, FCX, FP, FU, FX, FXU, GS, GU, I, Kl, LA, LQ, LS, MU, OV, PC, PCD, PDC, PF, PU, QRS, QSV, QU, QV, SC, SCTU, SD, SDU, SG, SMU, ST, STD, STGD, STU, SU, SX, TCG, TD, TDG, TDS, TS, TSD, TX, XB, XC, XCU, XD, XDC, XF, XFC, XFCU, XFU, XL, XS, XSC, XSCU, XT, XU} \\
         \hline
         \multicolumn{3}{p{0.9\columnwidth}}{The references indicate where the class or complex 
            archetype was labeled as such for the first time.}
      \end{tabular}
   \end{table}

\end{appendix}

\end{document}